\documentclass{emulateapj}
\shorttitle{Rossby wave instability in 3D disks}
\shortauthors{M-K. Lin}

\usepackage{amsmath}
\usepackage{url}
\usepackage{bm}
\usepackage{appendix}
\newcommand{\p}{\partial}
\newcommand{\lmax}{l_\mathrm{max}}
\newcommand{\sbar}{\bar{\sigma}} 
\newcommand{\imgi}{\mathrm{i}} 
\newcommand{\ii}{\mathrm{i}}
\newcommand{\C}{\mathcal{C}^\lambda}
\newcommand{\wop}[1]{\mathcal{V}^{(#1)}}
\newcommand{\avg}[1]{\langle{#1}\rangle}
\newcommand{\zhat}{\hat{z}}
\newcommand{\rhat}{\hat{r}}
\newcommand{\he}{\mathcal{H}}
\newcommand{\amp}{\mathcal{A}}
\newcommand{\dd}{\delta}
\newcommand{\real}{\operatorname{Re}}
\newcommand{\imag}{\operatorname{Im}}
 \newcommand{\sgn}{\operatorname{sgn}}

\begin{document}

\title{Rossby wave instability in locally isothermal and polytropic
  disks: three-dimensional linear calculations}

\author{Min-Kai Lin}
\affil{Canadian Institute for Theoretical Astrophysics,
60 St. George Street, Toronto, ON, M5S 3H8, Canada}
\email[Email: ]{mklin924@cita.utoronto.ca}
\begin{abstract}
Numerical calculations of the linear Rossby wave instability (RWI) in
global three-dimensional (3D) disks are presented. The linearized fluid equations
are solved for vertically stratified, radially structured disks
with either a locally isothermal or polytropic equation of state, 
 by decomposing  the vertical dependence of the perturbed 
  hydrodynamic quantities into Hermite and Gegenbauer polynomials,
  respectively.  It is confirmed that the RWI operates in 3D. 
 For perturbations with vertical dependence assumed above, there 
is little difference in growth rates between 3D and two-dimensional (2D)
calculations. Comparison between 2D and 3D solutions  of this
  type suggest the RWI is predominantly a 2D instability and that
three-dimensional effects, such as vertical motion, to be interpreted
as a perturbative consequence of  the dominant 2D flow. The vertical flow around 
co-rotation, where vortex-formation is expected, is examined. 
In locally isothermal disks the expected vortex center remains in
approximate vertical hydrostatic equilibrium. For polytropic disks
the vortex center has positive vertical velocity, whose magnitude
increases with decreasing polytropic index $n$. 
\end{abstract}

\section{Introduction}
Theoretical modeling of protoplanetary disks lead to complex
structures that are unlikely to be described by smooth radial profiles 
\citep{terquem08, armitage11}. However, radially structured disks may 
develop the Rossby wave instability \citep[RWI, ][]{lovelace99,li00}, 
which leads to vortex-formation in the nonlinear
 regime \citep{li01}. Thus, the RWI may play a role in the evolution 
of protoplanetary disks.

The disk RWI is a dynamical instability associated with the presence of 
extrema in the ratio of vorticity to surface density, or
vortensity\footnote{This quantity is modified by a factor involving the disk 
entropy, if the latter is not constant.}. The instability results from
wave coupling across such an extremum. Its physics is similar to the
Papaloizou-Pringle instability
\citep[PPI,][]{papaloizou84,papaloizou85,papaloizou87,goldreich86,narayan87} 
which operate in pressure-supported, thick tori. The RWI operates in
thin, centrifugally-supported disks with non-power law
rotation profiles, and is insensitive to radial boundary conditions.

The relevance of the RWI in protoplanetary disks has been 
demonstrated in two situations. Variability in the efficiency of
turbulent angular momentum transport by the magneto-rotational
instability \citep{balbus91} can result in
the existence of `dead zones' \citep{gammie96}, in which the turbulent
viscosity is small. The radial boundary between a
dead zone and the actively accreting region is prone to the RWI 
\citep{varniere06, lyra08, lyra09, crespe11}, with 
observable consequences \citep{regaly12}. In 
addition to hydrodynamic angular momentum transport, the RWI may also 
assist planet formation formation by concentrating solids into
anti-cyclonic vortices \citep{barge95}.       


Another origin of the RWI in protoplanetary disks, which
motivated this study, is disk-planet interaction
\citep{goldreich79,goldreich80}. A sufficiently massive planet leads
to gap opening \citep{lin86}, while low mass protoplanets may open
gaps provided the disk viscosity is sufficiently small \citep{muto10,dong11}.   
Vortensity jumps across planet-induced shocks lead to the
necessary disk profile for the RWI to develop around gap edges
\citep{koller03, li05,valborro07}. Subsequent vortex-formation
significantly affects disk-planet torques and migration \citep{ou07, 
  li09,yu10,lin10}.   

The above studies of the RWI have all employed the razor-thin disk
approximation (but note that the PPI was originally analyzed in
3D). \cite{yu09} have examined the RWI with a toroidal
magnetic field in a 3D but unstratified disk. \cite{meheut10} first 
demonstrated the RWI in nonlinear hydrodynamic simulations of 3D stratified
disks \citep[ later with improved resolution in][]{meheut12b}, while
\cite{umurhan10} analyzed the RWI in approximate 3D disk models. 

Recently, \cite{meheut12} calculated linear RWI modes in a 
three-dimensional, globally isothermal disk, which displayed
vertical motion. In this paper, we compute linear RWI modes in
three-dimensional disks across a range of parameter values, including
different equations of state. Our focus here is on how these affect
the vertical flow in the co-rotation region, where vortex-formation is
known to occur \citep{li01}. 


This paper is organized as follows. In \S\ref{eqm_disks} we list the
governing equations and describe our disk models. We derive the
linearized fluid equations in \S\ref{linearized} and describe our
numerical methods in \S\ref{numerics}. Results are presented in
\S\ref{isothermal} for locally isothermal disks and in
\S\ref{polytropic} for polytropic disks. In \S\ref{meheut} 
 we briefly examine a linear mode qualitatively different to those 
  above, found in a disk model involving $\kappa^2<0$ 
  \citep[taken from][]{meheut10}, where $\kappa$ is the epicycle
  frequency. 
We summarize and discuss
our results in  \S\ref{summary}, including limitations of our
calculations.  

\section{Governing equations, disk models and assumptions}\label{eqm_disks}
We consider a three-dimensional, inviscid, non-self-gravitating disk
orbiting a star of mass $M_*$ and adopt $(r,\phi,z)$ cylindrical polar
co-ordinates centered on the star. The frame is non-rotating. The
governing equations are the 3D Euler equations:   
\begin{align}
&\frac{\p\rho}{\p t} + \nabla\cdot(\rho\bm{v}) = 0, \\
&\frac{\p\bm{v}}{\p t} + \bm{v}\cdot\nabla\bm{v} =
  -\frac{1}{\rho}\nabla P -\nabla\Phi_*, \\
&P = P(r,\rho)\label{EOS},
\end{align}
where $\rho$ is the density, $P$ is the pressure, $\bm{v}$ is the
velocity field and $\Phi_*$ is the gravitational potential due to the
central star. Eq. \ref{EOS} is an equation of state (EOS),
specified later, such that the pressure may be calculated without an
energy equation. 

  We assume the disk is geometrically thin  
  so that $\Phi_*$ may be approximated as    
  \begin{align}\label{thin_disk_pot}
    \Phi_*(r,z) = -\frac{GM_*}{\sqrt{r^2 + z^2}}\simeq
    -\frac{GM_*}{r}\left(1-\frac{z^2}{2r^2}\right). 
  \end{align} 
  This approximation is adopted so that the resulting equilibrium
  density field has a convenient functional form suitable for the
  application of orthogonal polynomials (see \S\ref{numerics}). This
  greatly simplifies the numerical problem. Henceforth we use the
  approximate potential for self-consistency. 

The unperturbed disk is steady, axisymmetric with no meridional
velocity   ($\partial_t=\partial_\phi=v_r=v_z=0$).                                                    
The disk is stratified with $\rho=\rho(r,z)$ set by 
vertical hydrostatic balance. The  azimuthal velocity is $v_\phi=r\Omega$, 
where $\Omega$ is the angular speed. $v_\phi$ is set by radial balance between 
pressure, stellar gravity and centrifugal forces. Because the
disk is thin, the angular velocity is close to Keplerian, i.e.
$\Omega\simeq\Omega_k\equiv(GM_*/r^3)^{1/2}$.       

To introduce radial structure, we choose the  unperturbed surface
density profile to be   
\begin{align}\label{init_sigma}
  \Sigma(r) &\equiv \int^{\infty}_{-\infty}\rho dz \notag\\
  &=  \Sigma_0\left(\frac{r}{r_0}\right)^{-\alpha}
  \left\{1 + (\mathcal{A} - 1)\exp{\left[-\frac{(r-r_0)^2}{2\Delta
        r^2}\right]}\right\}   
\end{align}
\citep{li00}.  
Eq. \ref{init_sigma} corresponds to a Gaussian surface density bump
centered at $r=r_0$, width $\Delta r$ and amplitude $\mathcal{A}$, on
top of a background power-law profile with index $-\alpha$.  
Since disk self-gravity is ignored, the surface density scale
$\Sigma_0$ is arbitrary. 

To specify the three-dimensional disk structure, we choose the EOS to
be either locally isothermal or polytropic. These are described below.    

\subsection{Locally isothermal disks}
For locally isothermal disks the pressure is calculated as  
\begin{align} 
P = c_s^2(r)\rho,
\end{align}
where $c_s(r)$ is the sound-speed given by $c_s =H\Omega_k$ and $H(r)$ is
the disk scale-height. The unperturbed density is
\begin{align}\label{unperturbed}
  \rho(r,z) =
  \frac{\Sigma(r)}{\sqrt{2\pi}H(r)}\exp{\left[-\frac{z^2}{2H^2(r)}\right]}. 
\end{align} 
In the numerical calculations we will choose $H(r) = hr$ with $h$
being a constant aspect-ratio, since this is a typical model for
protoplanetary disks\footnote{This choice also enables us to compare the
locally isothermal disk with a polytropic disk with constant
aspect-ratio.}.   
The exponential decay means the gas density becomes 
negligible after a few scale heights. Thus the vertical domain  can be
taken to be $z\in[-\infty,\infty]$, even though we have made the
thin-disk approximation.

  \subsubsection{Approximate equilibrium}\label{iso_eqm}
   For simplicity, we set the azimuthal velocity to  
  \begin{align}\label{unperturbed_vphi}
    v_\phi^2 = \left.\frac{r}{\rho}\frac{\partial P}{\partial r}\right|_{z=0} 
    + r\left.\frac{\partial\Phi_*}{\partial r}\right|_{z=0}.
  \end{align}
  Away from the midplane the deviation from exact radial balance is
  proportional to $h^2\ll 1$ for a 
  thin disk \citep{tanaka02}.    
  We adopt Eq. \ref{unperturbed_vphi} to allow us to apply
  standard solution methods.  

  To test whether or not the precise form of $\Omega$ affect our
  results, we also considered setting $\rho\to\Sigma$ in
  Eq. \ref{unperturbed_vphi}, which gives the velocity profile
  $v_{\phi,\mathrm{2D}}$ for a razor-thin disk. For our fiducial
  calculation, growth rates differ by $\sim1\%$ between adopting  
  Eq. \ref{unperturbed_vphi} or $v_{\phi,\mathrm{2D}}$, and we
  observe the same flow structure. 

In fact, locally isothermal disks
generally have  differential rotation in $z$, i.e. $\Omega=\Omega(r,z)$, unless the
disk is also globally isothermal.  It is therefore important to note
that in assuming Eq. \ref{unperturbed_vphi} , we have artificially suppressed baroclinic
effects. We discuss some justification for this in \S\ref{caveats} and
 Appendix \ref{expressions}. Although the chosen basic state is not in
exact equilibrium, setting $\Omega=\Omega(r)$ greatly simplifies the
linear equations as the only vertical dependence of the basic state is through the
exponential factor in $\rho$.  It allows us to address the specific
question of whether or not vertical density stratification  has any
effect on the RWI, without the complication of baroclinic
instabilities \citep{knobloch86,umurhan12}.     

%

\subsection{Polytropic disks}
In order to set up a more self-consistent basic state, that is,
$\Omega=\Omega(r)$  and a finite vertical domain, we also consider
polytropic disks, for which 
\begin{align}
  P = K\rho^{1+\frac{1}{n}}, 
\end{align}
where $K$ is a constant and $n$ is the polytropic index. Vertical 
hydrostatic equilibrium imply 

\begin{align}\label{poly_den}
\rho(r,z) &= \left[ \frac{GM_*H^2(r)}{2K(1+n)r^3} \right]^n\left[1 -
  \frac{z^2}{H^2(r)} \right]^n\notag\\ 
          &\equiv \rho_0(r)\left[1 - \frac{z^2}{H^2(r)}\right]^n
 \end{align}
Here, $z=H$ is the disk surface where
$\rho(r,H)=0$. Thus, when discussing polytropic disks $H$ is referred 
to as the disk thickness. 

The function $H(r)$ and mid-plane density $\rho_0(r)$ are calculated
through  
\begin{align}
\Sigma(r) = \rho_0(r)H(r)I_n,
\end{align}
where $I_n\equiv\int^1_{-1}(1-x^2)^ndx$, with $\rho_0(r)$ related to $H(r)$ 
by Eq. \ref{poly_den} and $\Sigma(r)$ given by Eq. \ref{init_sigma}. We can therefore write
\begin{align}\label{poly_thick}
H(r) = H_0\left[\frac{\Sigma(r)}{\mathcal{A}\Sigma_0}\right]^{\frac{1}{2n+1}}
       \left(\frac{r}{r_0} \right)^{\frac{3n}{2n+1}},
\end{align}
where $H_0=H(r_0)$ is the disk thickness \emph{at the bump radius}. We
parametrize it by writing $H_0=hr_0$ so that $h$ is the aspect-ratio
at $r_0$. 
Note that a surface density enhancement by a factor $\mathcal{A}$
corresponds to an enhancement of the disk thickness  by a factor
$\mathcal{A}^{1/(2n+1)}$.  

For a polytropic disk the azimuthal velocity is strictly independent of $z$
\citep[e.g.][]{papaloizou84}. It is given
by  \begin{align}\label{rot_poly} v_\phi^2(r) =
  r\frac{\partial}{\partial r}\Phi_*(r,H)  =
  \frac{GM_*}{r}\left(1 - \frac{3H^2}{2r^2} +\frac{H}{r}\frac{dH}{dr}
  \right), \end{align} where the second equality follows from the
approximation for the stellar potential in a  thin disk
(Eq. \ref{thin_disk_pot}).  

 Of course, given
  $H(r)$ one can obtain the azimuthal velocity  $v_{\phi,e}$
  corresponding to the exact gravitational potential of a point
  mass. For our fiducial setup, the difference in growth rate is
  $<4\%$ between using $v_{\phi,e}$ and using $v_\phi$ above, and we
  observe no difference in flow structure. However, we will use
  $v_\phi$ so that the equilibrium density and velocity fields are
  self-consistent and in exact balance with the same 
  potential.

\section{Linearized equations}\label{linearized}
In this section we derive the governing equation for small
disturbances in the disk. As described above, the basic state is
$\rho=\rho(r,z)$ and $\bm{v}=(0,r\Omega,0)$, with $\Omega=\Omega(r)$. 
The perturbed state is assumed to have the form   
\begin{align}
  \rho&\to\rho + \real[\dd\rho(r,z)\exp{\imgi(\sigma t + m\phi)}],\\
  P&\to P + \real[\dd P(r,z)\exp{\imgi(\sigma t + m\phi)}]  ,\\ 
  \bm{v}&\to\bm{v}+\real[\dd\bm{v}(r,z)\exp{\imgi(\sigma t +
      m\phi)}],  
\end{align}
where $\sigma = \sigma_R + \ii\gamma$  is a complex frequency
($\sigma_R,\,\gamma$ being real) and $m$ is the azimuthal wave-number
taken to be a positive integer. We will omit writing `$\real$' below, with
the understanding that physical solutions correspond to real parts of
the complex perturbations.      

For the locally isothermal equation of state, the linearized momentum
equations give 
\begin{align}
&\dd v_r = -\frac{\ii c_s^2}{D}\left( \sbar \frac{\p W}{\p r}  + \frac{2m\Omega W}{r}\right),\\
&\dd v_\phi = \frac{c_s^2}{D}\left( \frac{\kappa^2}{2\Omega} \frac{\p
    W}{\p r} + \frac{\sbar m W}{r} \right),\\ 
&\dd v_z  = \frac{\ii c_s^2}{\sbar}\frac{\p W}{\p z}\label{dvz_expression},
\end{align}
where $W\equiv \delta\rho/\rho$ is the relative density perturbation, 
$\sbar \equiv \sigma + m\Omega(r)$ is the shifted frequency, 
$D \equiv \kappa^2 - \sbar^2 $, and 
\begin{align}\label{ksq}
\kappa^2 = \frac{1}{r^3}\frac{\p}{\p r}\left(r^4\Omega^2\right)
\end{align}
is the square of the epicycle frequency. Corresponding equations 
for the polytropic disk are very similar, and are readily obtained by
setting $c_s$ to unity and replacing $W\to S\equiv \dd P/\rho$ where
$S$ is the enthalpy perturbation. 

Inserting the perturbed velocity field 
into the linearized continuity equation
\begin{align}
  \ii\sbar\dd\rho + \frac{1}{r}\frac{\p}{\p r}\left(r\rho\dd v_r\right)
  + \frac{\ii m}{r}\rho\dd v_\phi + \frac{\p}{\p z}\left(\rho \dd v_z\right) = 0,
\end{align}
yields, for locally isothermal disks:
\begin{align}\label{3d_linear}
  r\dd\rho =& \frac{\p}{\p r}\left(\frac{r\rho c_s^2}{D} \frac{\p W}{\p r}\right) + \frac{2mW}{\sbar}
  \frac{\p}{\p r}\left(\frac{c_s^2\rho\Omega}{D}\right) \notag\\
&- \left(\frac{m^2 c_s^2\rho}{rD}\right) W 
  -\frac{rc_s^2}{\sbar^2}\frac{\p}{\p z}\left(\rho\frac{\p W}{\p
    z}\right), 
\end{align}
and for polytropic disks: 
\begin{align}\label{3d_linear_poly}
r\dd\rho =&  \frac{\p}{\p r}\left(\frac{r\rho }{D} \frac{\p S}{\p r}\right) + \frac{2mS}{\sbar}
  \frac{\p}{\p r}\left(\frac{\rho\Omega}{D}\right) \notag\\
&- \left(\frac{m^2\rho}{rD}\right) S
  -\frac{r }{\sbar^2}\frac{\p}{\p z}\left(\rho\frac{\p S}{\p
    z}\right). 
\end{align}

We remark that Eq. \ref{3d_linear} is in fact valid for locally 
isothermal disks with any fixed sound-speed profile $c_s(r)$, 
assuming the equilibrium azimuthal velocity is independent of $z$ 
  (Appendix \ref{expressions}).  
Also note that 
Eq. \ref{3d_linear_poly} is actually valid for any barotropic EOS
, i.e. whenever $P=P(\rho)$. The 3D problem is to solve 
Eq. \ref{3d_linear}---\ref{3d_linear_poly}, which will
generally describe  disturbances depending on $(r,\phi,z)$ and motion 
in all three directions.

\subsection{Relation to the two-dimensional problem}
We define the 2D problem as solving
Eq. \ref{3d_linear}---\ref{3d_linear_poly} subject to
$\partial_z=0$. 
Denoting the corresponding solutions as
$W_\mathrm{2D}(r),\,S_\mathrm{2D}(r)$ and inserting them into the
governing equations yields, after vertical 
integration, 
\begin{align}\label{3d_linear_2D}
  r\dd\Sigma =& \frac{d}{d r}\left(\frac{r\Sigma c_s^2}{D} \frac{d W_\mathrm{2D}}{d r}\right) + \frac{2mW_\mathrm{2D}}{\sbar}
  \frac{d}{d r}\left(\frac{c_s^2\Sigma\Omega}{D}\right) \notag\\
&- \left(\frac{m^2 c_s^2}{rD}\right)\Sigma W_\mathrm{2D}
\end{align}
for locally isothermal disks and
\begin{align}\label{3d_linear_poly_2D}
r\dd\Sigma =&  \frac{d}{d r}\left(\frac{r\Sigma }{D} \frac{d S_\mathrm{2D}}{d r}\right) + \frac{2mS_\mathrm{2D}}{\sbar}
  \frac{d}{d r}\left(\frac{\Sigma\Omega}{D}\right) \notag\\
&- \left(\frac{m^2\Sigma}{rD}\right) S_\mathrm{2D}
\end{align}
for polytropic disks, where $\dd\Sigma =
\int^{\infty}_{-\infty}\dd\rho dz$ is the surface density
perturbation. Note that $W_\mathrm{2D}=\delta\Sigma/\Sigma$ is the
relative surface density perturbation, and
$S_\mathrm{2D}=\delta\Pi/\Sigma$ where $\delta\Pi$ is the perturbation
to the vertically integrated pressure ($\Pi
=\int^{\infty}_{-\infty} P dz $). Solutions to 
Eq. \ref{3d_linear_2D}---\ref{3d_linear_poly_2D} describe disturbances
which only depend on $(r,\phi)$ and there is no vertical motion.  

As defined here, the 2D problem and 3D problem involves the  same
background disk, which is three-dimensional. However, the
governing equation for linear disturbances in razor-thin disks have
the same form as  Eq. \ref{3d_linear_2D}---\ref{3d_linear_poly_2D}
when the razor-thin disk has a locally isothermal or barotropic EOS in
the form $\Pi = c_s^2(r)\Sigma$ or $\Pi=\Pi(\Sigma)$, respectively.

\subsection{Co-rotation singularity and the
  RWI}\label{corot_singularity} 

Inspection of the 2D equations, Eq.
\ref{3d_linear_2D}---\ref{3d_linear_poly_2D}, reveal a potential 
singularity when 
$\sbar(r_c)=0$, where $r_c$ is the co-rotation radius defined by    
\begin{align}\label{rc} 
 \sigma_R + m\Omega(r_c) = 0. 
\end{align}
This co-rotation singularity  can be rendered ineffective if
$r_c$ also satisfies  
\begin{align}
  &\frac{d}{dr}\left.\left(\frac{c_s^2}{\eta}\right)\right|_{r_c} = 0 
  \quad \text{for locally isothermal disks}\label{vortensity_extreme1},\\
  &\frac{d}{dr}\left.\left(\frac{1}{\eta}\right)\right|_{r_c} = 0 
   \,\,\quad \text{for polytropic disks}\label{vortensity_extreme2},
\end{align}
where 
\begin{align}
\eta \equiv \frac{\kappa^2}{2\Omega\Sigma}
\end{align}  
is the vortensity. The quantity $\eta/c_s^2$ can be seen as a 
generalized vortensity \citep{li00}, but for convenience we will simply
use `vortensity' in the discussion below. 
Thus there can exist 2D neutral disturbances with co-rotation at a
vortensity extremum, for which the 2D linear operator is real and
regular everywhere.     


Strictly speaking, co-rotation singularities only concern neutral 
disturbances ($\gamma = 0$). In practice we are interested in growing
solutions ($\gamma < 0$) so such singularities do not arise in the numerical 
computation. Nevertheless, the discussion above is important because 
the growth rates we find are typically
$|\gamma|\ll\Omega(r_0)$. Furthermore, association of $r_c$ with a 
vortensity extremum forms the basis of 
the RWI.  

In studies employing razor-thin disks, the RWI has 
largest disturbance amplitude in the co-rotation region where 
$|\sbar^2|\ll\kappa^2$. It can be shown that such modes can only be unstable if there 
exists vortensity extrema in the disk \citep[e.g.][]{lin10}. Indeed, the RWI
is found to have with co-rotation radius $r_c$ close to a vortensity
minimum \citep{lovelace99,li00,lin11a}. 

It is precisely linear modes with the above properties which we wish
to explore 
in 3D. However, we do not expect such modes to have  
significant $z$-dependence in their relative density or enthalpy 
perturbation around co-rotation. 
From the linearized vertical equation of 
motion we see that
\begin{align*}   
  \dd v_z \propto \frac{1}{\sbar}\frac{\p X}{\p z},
\end{align*}
where $X$ is $W$ or $S$ depending on the EOS. Near co-rotation 
where $|\sbar|$ is small, $|\p_zX|$ should be almost negligible. 
Otherwise, even small vertical gradients in density or enthalpy
perturbation will cause significant vertical motion, and linearization
becomes invalid.  



\section{Numerical procedure}\label{numerics}

In principle one could attempt a numerical solution to the partial
differential equations (PDE) above, for example by finite-differencing
in the $(r,z)$ plane. However, since one of our goals is to assess 
three-dimensional effects, it is more useful to have a numerical
scheme that automatically  separates out the 2D problem from the full
3D problem.  

We begin by making the co-ordinate transformation
\begin{align}
(\hat{r},\hat{z})&\equiv(r, z/H),\\
\left(\frac{\p}{\p r}, \frac{\p}{\p z}\right)
&= \left( \frac{\p}{\p \hat{r}} - \zhat\frac{H^\prime}{H}\frac{\p}{\p\zhat}, 
 \frac{1}{H}\frac{\p}{\p\zhat}\right), 
\end{align}
where $^\prime$ denotes differentiation with respect to the argument. 
In this co-ordinate system the background density is separable,
i.e. $\rho(\hat{r},\zhat) = g(\hat{r})f(\zhat)$, where
$f=\exp{(-\zhat^2/2)}$ for locally  isothermal disks and
$f=(1-\zhat^2)^n$ for polytropic disks. This motivates us to seek
solutions of the form  
\begin{align}
  W &= \sum_{l=0}^{\infty} W_l(\hat{r})\he_l(\zhat),\label{hermite_expand} \\
  S &= \sum_{l=0}^{\infty} S_l(\hat{r})\C_l(\zhat), \label{gegen_expand}
\end{align}
where $\he_l$ is a Hermite polynomial of order $l$ and $\C_l$ is
a Gegenbauer polynomial of index $\lambda$ and order $l$.  Note
  that radial and vertical variations are coupled because
  $\hat{z}=\hat{z}(r)$ through $H(r)$.

These 
polynomials satisfy the orthogonality relations
\begin{align}
&  \int_{-\infty}^\infty \he_k(\zhat) \he_l(\zhat)\exp{(-\zhat^2/2)}d\zhat = \sqrt{2\pi} l! \delta_{kl},\\   
&\int_{-1}^1 \C_k(\zhat) \C_l(\zhat)(1-\zhat^2)^{\lambda - 1/2}d\zhat 
= \frac{\pi 2^{1-2\lambda}\Gamma(l+2\lambda)}{l!(l+\lambda)\Gamma^2(\lambda)}\delta_{kl},
\end{align}
where $\delta_{kl}$ here is the Kronecker delta and $\Gamma$ is the Gamma 
function \citep{stegun65}. For polytropic disks, we choose the
parameter $\lambda$ to be  
\begin{align}
\lambda = n - \frac{1}{2}. 
\end{align}
Consequently, for a polytropic  index $n=1.5$, $\mathcal{C}^{1}_l$ are 
the Chebyshev polynomials of the second kind, and for $n=1$,
$\mathcal{C}^{1/2}_l$  are the Legendre polynomials. Eigenfunction
expansions in $\hat{z}$ is a standard method to account
for vertical dependence in disk problems \cite[e.g.][]{okazaki85,
  papaloizou85,takeuchi98,tanaka02}.

It is important to keep in mind that by assuming the above
  decompositions (Eq. \ref{hermite_expand}---\ref{gegen_expand}) we
  restrict the type of perturbations to those satisfying certain
  physical conditions implied by the orthogonal polynomials at the
  upper disk boundary. In the locally isothermal disk we require the
  kinetic energy density to be bounded at large heights \citep{takeuchi98},
  and for polytropic disks a regularity condition applies at $\hat{z}=\pm1$
  \citep{papaloizou85}. Such 
  perturbations can be decomposed as above because the polynomials form a
  complete set \citep{zhang06}. On the other hand, the above specific
  decomposition cannot be applied if one considers other vertical
  boundary conditions (e.g., conditions imposed at other
  heights).

After transforming the governing equations into $(\rhat,\zhat)$
co-ordinates, we insert the anstaz
Eq. \ref{hermite_expand}---\ref{gegen_expand} into
Eq. \ref{3d_linear}---Eq. \ref{3d_linear_poly}, multiply
by $\he_k,$ and $ \C_k$ respectively, then integrate vertically. This
procedure yields an equation of the form    

\begin{align}\label{operators}
  A_l X_l + B_l X_{l-2} + C_l X_{l+2} = 0,
\end{align}
where $X_l$ is $W_l$ or $S_l$, and  $A_l,\,B_l,\,C_l$ are linear
operators  which only depend on $r$ and $\sigma$, but are different
for the two EOS (see Appendix \ref{expressions}). For each $l$ there
is a separate equation with the operators $B_l,\,C_l$ representing
coupling with the $l\pm2$ modes. Note that
$B_l$ is set to zero when $l=0,\,1$.  

We have now transformed the governing partial differential equation into an infinite set of coupled 
ordinary differential equations (ODE). In practice we truncate the solution at $\lmax$, i.e. $X_l\equiv 0$ 
for $l>\lmax$. The decomposition has the advantage that for modes
nearly independent of $z$, $\lmax$ can be small. In the simplest
case of setting $\lmax=0$, we only solve
\begin{align}    
A_0X_0 = 0,
\end{align}
which is the 2D problem. That is, if $\lmax=0$ then
$W_0=W_\mathrm{2D}$  and $S_0=S_\mathrm{2D}$.

\subsection{Matrix methods}
We now proceed to a numerical solution to the linear problem. 
We discretize the linear operators and solutions on 
a grid which divides the radial range $r\in[r_i, r_o]$  into 
$N_r$ uniformly spaced points. The coupled
set of ODEs then become a single matrix equation. 
This is denoted generically as 
\begin{align}\label{matrix_eq}
  \bm{M}\bm{x} = \bm{0},
\end{align}
where the square matrix 
$\bm{M}$ represents the discretized linear operator and the vector $\bm{x}$ 
is the discretized solution. The size of the matrix
and vector depends on $\lmax$. For example 
setting $\lmax = 4$, Eq. \ref{matrix_eq} then represents 
the discretized version of   
\begin{align*}
A_0X_0 + C_0X_2 \phantom{+1C_2X_4}  &= 0, \\
B_2X_0 + A_2X_2 + C_2X_4 &= 0, \\
B_4X_2 + A_4X_4  &= 0, 
\end{align*}
for which $\bm{M}$ is a $3N_r\times3N_r$ matrix 
and $\bm{x}$ is a vector of length $3N_r$.


The matrix problem, Eq. \ref{matrix_eq}, is a set of homogeneous linear
equations. Non-trivial solutions 
exist if 
\begin{align}
\mathrm{det} \bm{M}= 0.
\end{align}
The complex frequency $\sigma$ is required to be such that the matrix
$\bm{M}(\sigma)$  is singular. We have used two approaches to achieve
this. The first is to consider the usual eigenvalue problem: 
\begin{align}\label{standard_eigen}
  \bm{M}(\sigma)\bm{x} = \nu \bm{x}. 
\end{align}
Starting with a trial $\sigma$, standard matrix software\footnote{
We used \texttt{LAPACK}.} may be used to find the eigenvalues  
$\nu$ and associated eigenvectors $\bm{x}$. We then apply
Newton-Raphson iteration to solve
$\nu_\mathrm{min}/|\nu_\mathrm{max}|=0$ by varying $\sigma$,  where
$\nu_\mathrm{min, max}$ corresponds to eigenvalues of smallest and
largest absolute value found from Eq. \ref{standard_eigen}.   

Another approach is to perform a singular value
decomposition\footnote{We used \texttt{LAPACK} for 
a direct decomposition. We also performed the SVD with 
 \texttt{PROPACK} (available at
\url{http://soi.stanford.edu/$\sim$rmunk/PROPACK/}), which is an
iterative method. These gave the same results.} 
(SVD) of $\bm{M}$, 
so that 
\begin{align}
\bm{M} = \bm{U}\mathrm{diag}(s_1, s_2, ... )\bm{V}^\dagger,
\end{align}
where $\bm{U},\,\bm{V}$ are unitary matrices ($^\dagger$ denotes 
 Hermitian conjugate) and the real numbers $s_i\geq 0$ are 
the singular values of $\bm{M}$. 
The columns of $\bm{V}$ are the 
right singular vectors of $\bm{M}$. If $\mathrm{min}(s_i)=0$  
then $\bm{M}\bm{x}_0=\bm{0}$, where $\bm{x}_0$
is the right singular vector associated with $\mathrm{min}(s_i)$. 
We therefore use Newton-Raphson iteration to zero the quantity 
$F\equiv \bm{x}_0^\dagger \bm{M} \bm{x}_0/\bm{x}_0^\dagger \bm{x}_0$
by varying $\sigma$.  

These methods give the same result.  
We always perform the SVD for the final matrix $\bm{M}(\sigma)$ in
order to evaluate $R^{-1}$, where $R\equiv
\mathrm{max}(s_i)/\mathrm{min}(s_i)$ is the condition number of
$\bm{M}$. Since $R=\infty$ for a singular matrix, we only accept
solutions for which $R^{-1}$ is zero at machine precision (typically
$R^{-1}\lesssim10^{-15}$).  The matrix methods outlined above was
also used in \cite{lin11a,lin11b}.

\subsection{Radial boundary conditions}\label{radial_bc}
For simplicity we impose $dX_l/dr=0$ at $r=r_i,\,r_o$. 
The RWI is associated with internal structure away from
boundaries. Consequently, it is insensitive to 
radial boundary conditions in razor-thin disks \citep{valborro07,
  lin11a}. We assume this still holds in 3D.  For example, 
approximate 3D disk models developed by \cite{umurhan08, umurhan10}, 
in which the inner/outer disk boundaries play no role, also support
the RWI. 

As a check, additional calculations were performed
with: $\partial_r X =0$ applied at boundaries (which introduces     
mode coupling), different $r_i,\, r_o$ and a numerical
condition where boundary derivatives are approximated by interior
points. The last case is strictly a numerical procedure to generate a 
closed set of equations to solve. For the solutions of interest, these
experiments gave results with no appreciable difference.


\subsection{Fiducial setup}
We work in units such that $G=M_*=1$. Our standard disk spans
$r\in[r_i,r_o]=[0.4, 1.6]$ and has a surface density profile with
$\alpha =0.5$. The bump is located at $r_0=1$ with width parameter
$\beta=0.05$.  
We use $N_r=512$ grid points and first solve 
the 2D problem ($\lmax=0$), then use the obtained eigenvalue to 
start the iteration for the 3D problem, for which $\lmax=6$. We only
consider even $l$.

In  \S\ref{meheut} we will use the setup employed by 
\cite{meheut10} to examine a 3D RWI mode with $\kappa^2<0$ at 
the bump radius. This mode appears quite different 
to our standard setup with $\kappa^2>0$ everywhere.

\subsection{Results analysis}
The solution to the linear problem gives the complex radial functions     
$X_l$, which can be used to reconstruct the complex amplitudes, 
e.g. $\dd v_z(r,z)$ by using
Eq. \ref{hermite_expand}---\ref{gegen_expand} and 
Eq. \ref{dvz_expression}, but we are interested in physical (real)
solutions. We will often visualize the solution for a specific $m$ with     
two-dimensional plots. We explain below how these are obtained.   

The real perturbation is, e.g. $\real[\dd v_z \exp 
  \ii(\sigma t + m\phi)$, so the spatial dependence of a 
  physical perturbation is  
  \begin{align} 
    \dd v_z \to \real[\dd v_z]\cos{(m\phi)} - \imag[\dd
      v_z]\sin{(m\phi)}\label{real_pert}, 
  \end{align}
  and similarly for other variables. We focus on the solution near 
  the \emph{vortex core}, defined to be at $(r,\phi)=(r_0,\phi_0)$,
  where   
  \begin{align}
    &\cos{(m\phi_0)} =\phantom{-} \real[X(r_0,0)]/|X(r_0,0)|, \notag\\
    &\sin{(m\phi_0)} =-\imag[X(r_0,0)]/|X(r_0,0)|. 
  \end{align}
  The magnitude of the (real) perturbation is 
  arbitrary but its sign is fixed, e.g. $X$, 
  now representing the real density or enthalpy perturbation, is
  positive at $(r,\phi,z)=(r_0,\phi_0,0)$. In practice the vortex core
  is near a maximum midplane over-density.  

  We visualize results in the $(r,z)$ plane by setting
  $\phi=\phi_0$ in Eq. \ref{real_pert}. Similarly, perturbations are
  visualized in the $(r,\phi)$ plane at a chosen $z$, and in the 
  $(\phi,z)$ plane at $r=r_0$ with the azimuthal range set to  
  $\phi\in[\phi_0-\pi/2m,\phi_0+\pi/2m]$. For convenience we also
  define $\Omega_0\equiv\Omega(r_0)$ and $\kappa_0\equiv\kappa(r_0)$.

\section{Results: locally isothermal disks}\label{isothermal}
\begin{figure}[!t]
   \centering
   \includegraphics[width = 1.0 \linewidth]{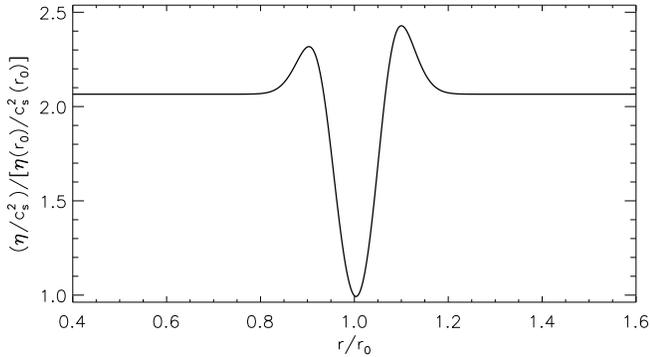}
   \caption{Background profile of the fiducial locally isothermal
     disk with $\amp=1.25$ and $h=0.07$, in terms of the generalized
     vortensity $\eta/c_s^2$, scaled by its value at the
     bump. Unstable modes are associated with the minimum at $r_0$. 
     \label{basic_iso}}
 \end{figure}

For locally isothermal disks we choose $h=0.07$ and 
$\mathcal{A}=1.25$ as a fiducial case. 
Recall $c_s^2\propto 1/r$, so that far from $r_0$ the generalized  
vortensity $\eta/c_s^2$ is flat, and is a minimum at $r_0$.  
The background disk is shown in Fig. \ref{basic_iso}. 
Note that $\kappa^2>0$ everywhere, and 
$\mathrm{min}(\kappa^2/\Omega_k^2)\simeq0.59$. 

 Recall that for locally isothermal disks we assumed an
  approximate basic state (\S\ref{iso_eqm}). The extent of inexact
  radial balance in the background depends on $h$
  \citep{tanaka02}. In a nonlinear simulation this 
  may lead to radial motion. To keep this effect
  fixed in comparing different linear calculations below, in this 
  section we fix $h$.

\subsection{Solution example}
We solved the fiducial case for $m\in[1,6]$. 
Table \ref{iso_freq} compares the eigenfrequencies obtained from the   
2D and 3D problems. Growth rates in 2D and 3D are very similar, so 
the instability is largely associated with $W_0$. We thus expect the
RWI to grow in 3D disks on the same time-scales as in the razor-thin
disks\footnote{This statement assumes the 2D problem give similar growth rates 
to the equivalent razor-thin disk setup, which we have checked to be
the case.} \citep[e.g.][]{li00}. 
The growth rate for the most unstable mode ($m=3$) is only $\simeq
0.06\Omega_0$ but this corresponds to $\sim 3$ orbits at $r_0$,  
so the instability operates on dynamical time-scales.

\begin{deluxetable}{ccc}
\tablecaption{Eigenfrequencies in the locally
  isothermal disk with $h=0.07$\label{iso_freq}}
\tablehead{\colhead{$m$} & \colhead{$-\sigma_R/(m\Omega_0)$} &
  \colhead{$-10^2\gamma/\Omega_0$} \\ 
\colhead{} & \colhead{} & \colhead{} }
\startdata
1 & 0.9960 (0.9960) & 2.8038 (2.8044) \\

2 & 0.9960 (0.9960) & 4.8931 (4.8985) \\

3 & 0.9961 (0.9960) & 5.7205 (5.7365) \\

4 & 0.9964 (0.9964) & 5.1245 (5.1843)\\

5 & 0.9972 (0.9971) & 3.4557  (3.5720) \\

6 & 0.9980 (0.9978)  & 1.8317 (1.9615)  \\
\enddata
\tablecomments{
  Values in brackets were obtained from the 2D problem. }
\end{deluxetable}



\begin{figure}[!t]
   \centering
   \includegraphics[scale=.425,clip=true,trim=0cm 1.8cm 0cm 0cm]{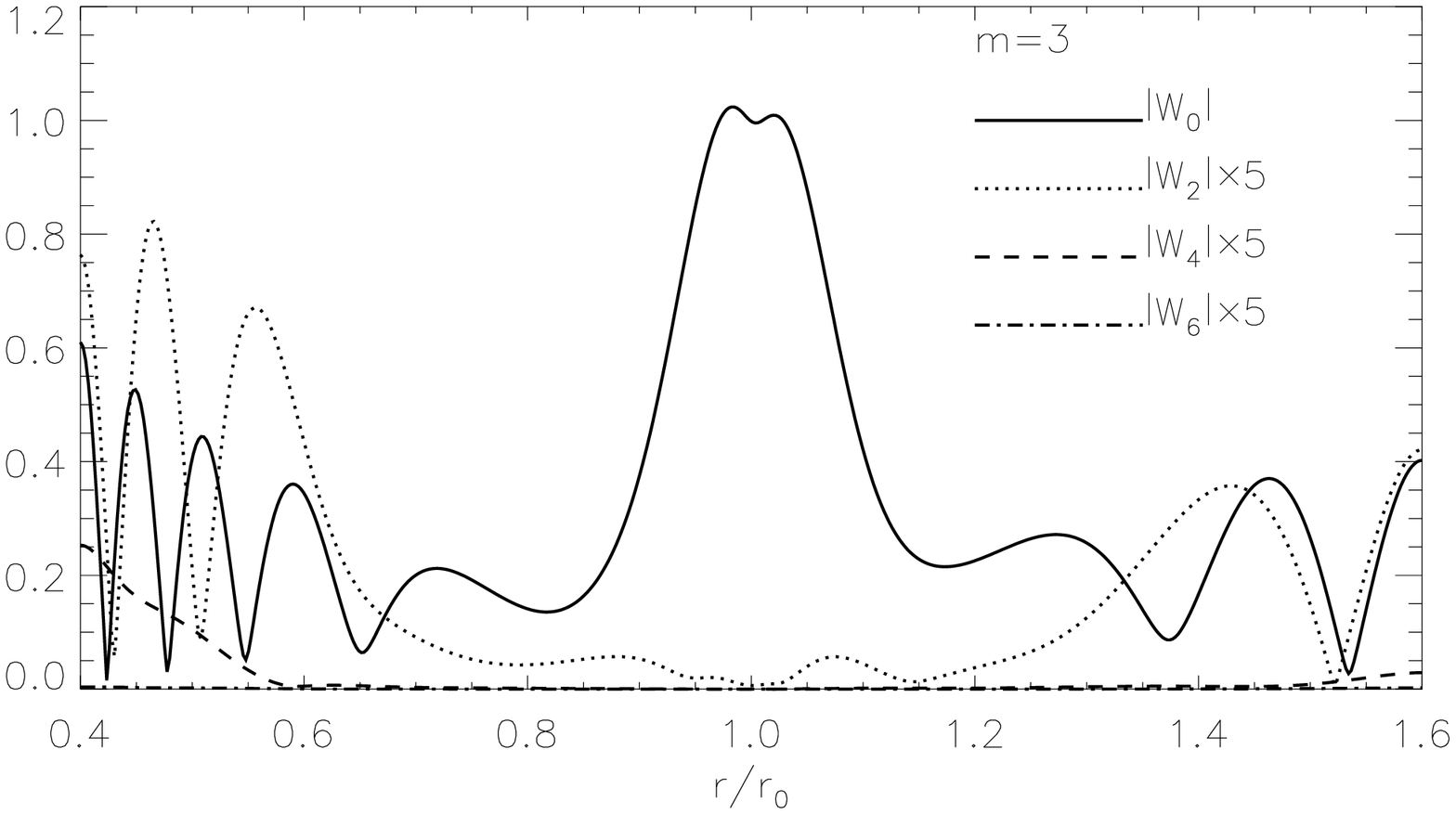}
   \includegraphics[scale=.425,clip=true,trim=0cm 0.0cm 0cm 0.2cm]{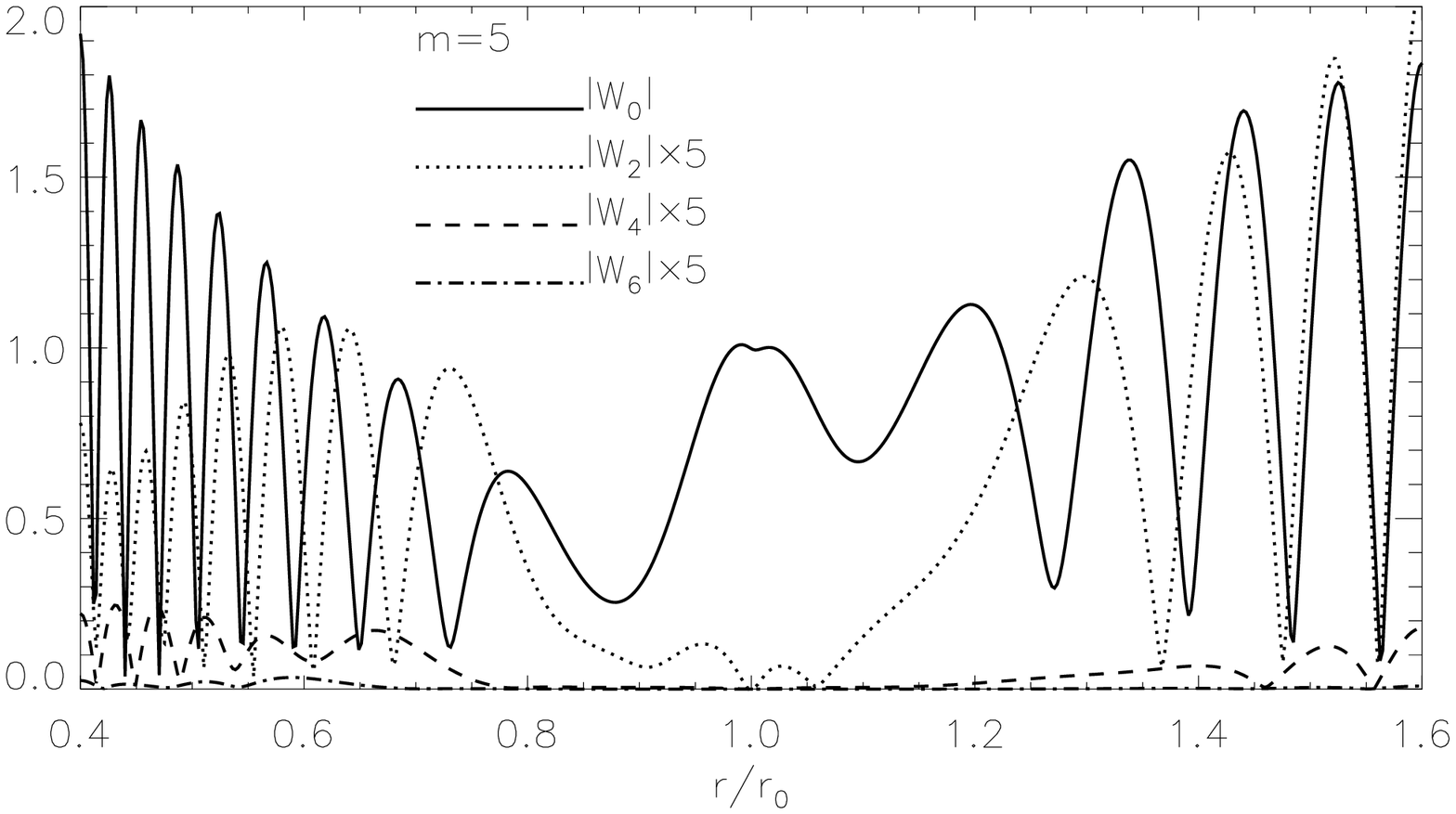} 
   \caption{Radial eigenfunctions $W_l$ for locally isothermal disk with $h=0.07$. 
    These are normalized by $|W_0(r_0)|$. The $l>0$ modes have also been 
    magnified in order to compare its radial structure with $W_0$. 
     \label{2d_3d_func}}
 \end{figure}

Fig. \ref{2d_3d_func} compares the radial functions $W_l$ for the 
$m=3$ and $m=5$ modes. In both cases, $W_0$ dominates over $W_{l>0}$, 
implying that the relative density perturbation is nearly 
$z$-independent. For $m=3$, $W_0$ itself is dominated by the
co-rotation region $r\sim r_0$, but for $m=5$ the amplitude in the
oscillatory region is larger than that around $r_0$. The $W_{l>2}$
modes are negligible, so three-dimensional effects are due to 
$W_2$. Unlike $W_0$, in both cases $|W_2|$ has largest amplitudes in the wave-like
regions towards the boundaries, and is smallest
near $ r_0$. This is consistent with the absorption of waves with $l>0$  
at co-rotation discussed in \cite{li03}.


It is well known that in the razor-thin disk,  as $m$ is increased the
RWI becomes more wave-like (as seen here for $W_0$) and is eventually
quenched \citep{li00}. This might contribute to the slightly smaller
growth rates obtained in the 3D problem than in the 2D problem (Table 
\ref{iso_freq}) since $W_{l>0}$ are 
wave-like (in addition to wave absorption at co-rotation). However,
this effect is unimportant because their amplitudes are much smaller
than $W_0$.    

Although the observed stabilization effect increases with $m$, $W_0$
loses its RWI character at high $m$. Thus, it can be said
that the RWI, considered as a low $m$, radially confined
non-axisymmetric  disturbance, 
has a growth rate determined by the 2D problem.     


\subsection{Flow in the $(r,\phi)$ plane}
\begin{figure}[!t]
   \centering
   \includegraphics[scale=.425,clip=true,trim=0cm 0.0cm 0cm
     0cm]{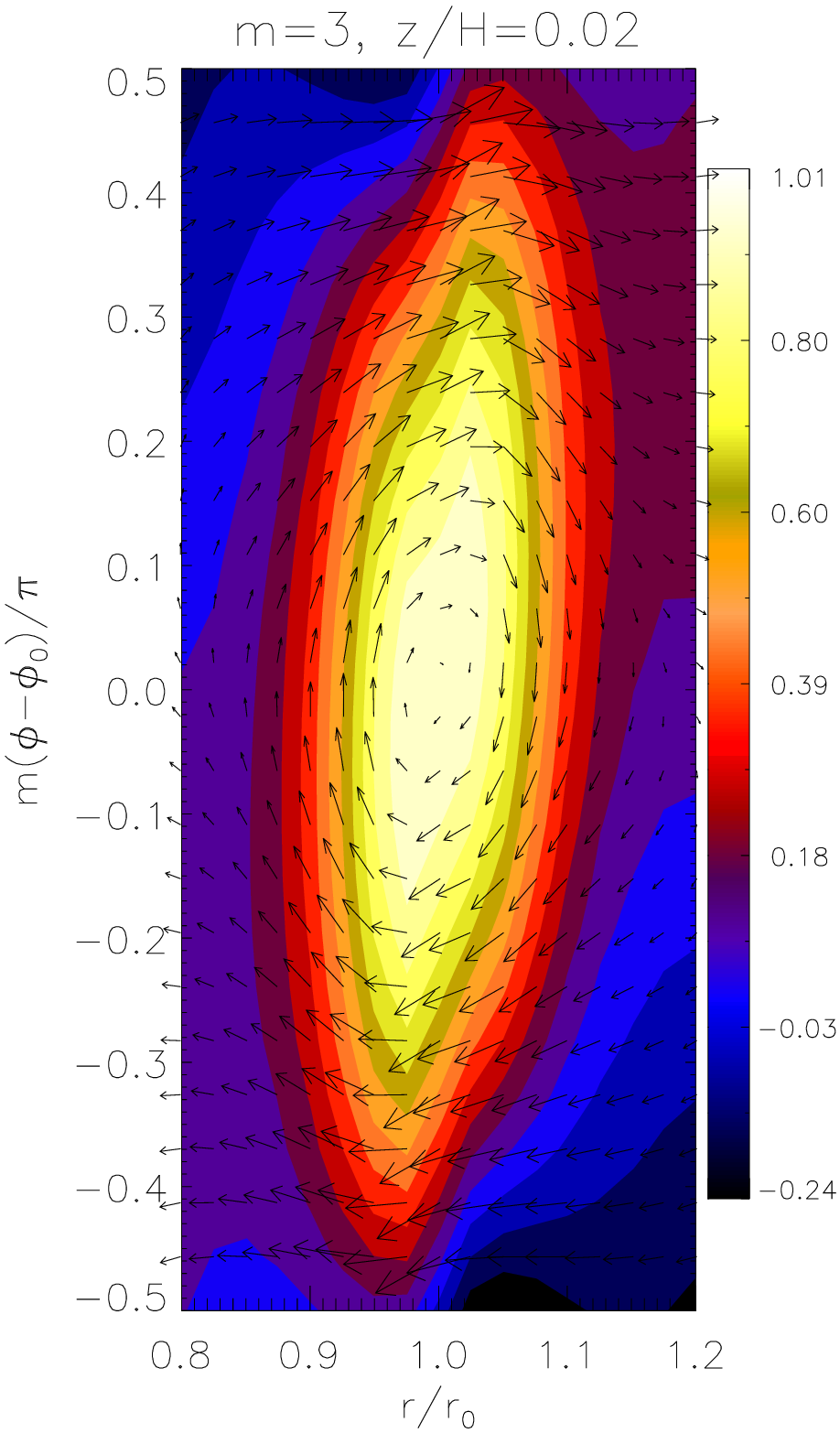}\includegraphics[scale=.425,clip=true,trim=2.2cm
     0.0cm 0cm 0.0cm]{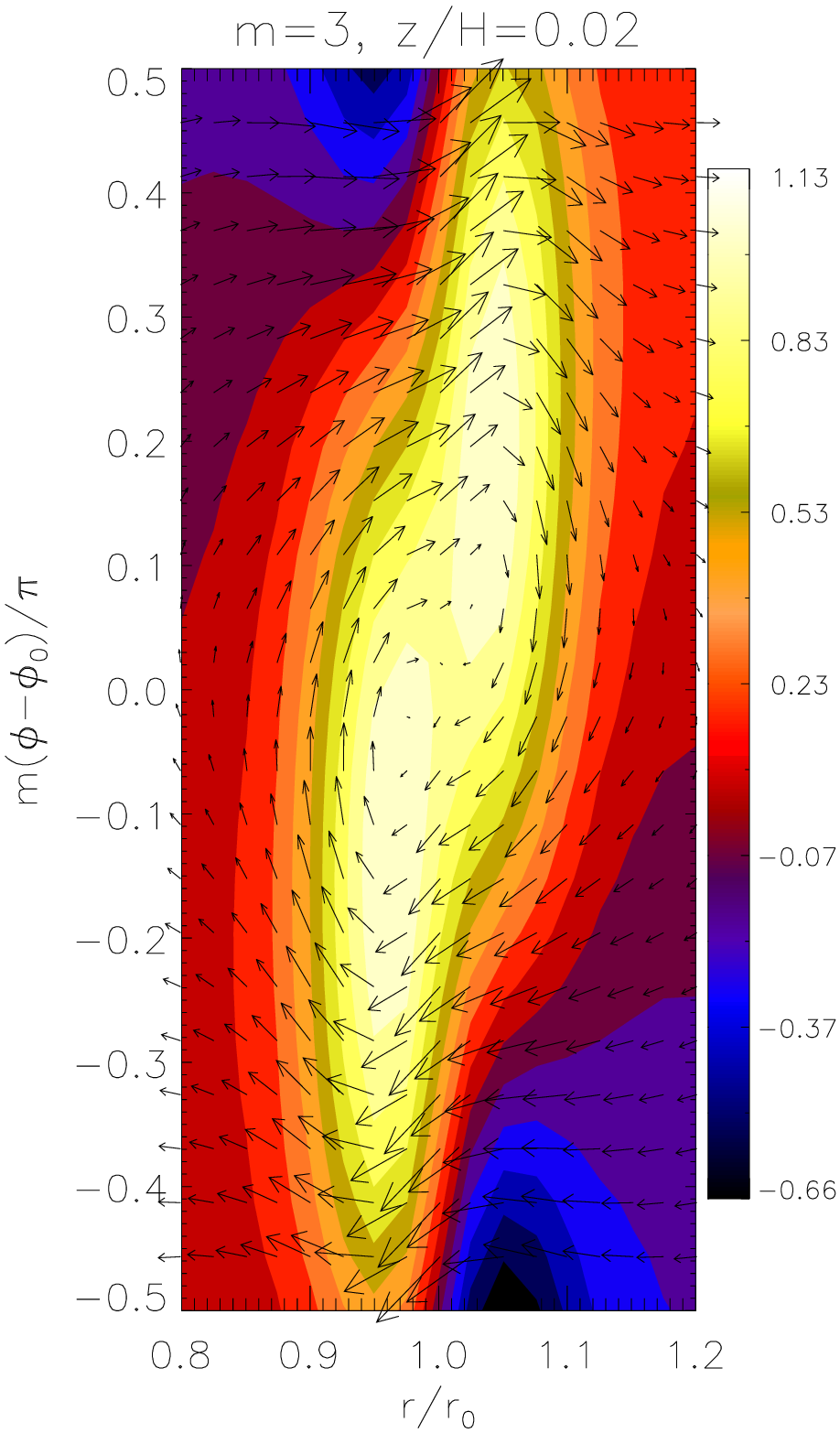} 
   \caption{Perturbed horizontal velocity field in the $(r,\phi)$
     plane, for the locally isothermal disk with $h=0.07$ and
     $\amp=1.25$ (left) 
      and
       with $\amp=1.6$ (right). The contours indicate relative density
       perturbation. The case with $\amp=1.6$ displays a double-peak in
       density perturbation, which is explained in \cite{li00}.
   \label{2dflow}}
 \end{figure}

Fig. \ref{2dflow} shows the perturbed velocity field in the $(r,\phi)$  
plane for the $m=3$ mode above, and  for a case with $\amp=1.6$
  (growth rate $\sim0.15\Omega_0$). The figure shows that
anti-cyclonic motion at an over-density is a 
robust feature. This confirms that the unstable modes found here are indeed the analog
of the RWI in razor-thin disks. The perturbed horizontal velocity
$(\dd v_r,\dd v_\phi)$  
has negligible variation with respect to $z$. 

\subsection{Vertical flow}
We now examine vertical flow associated with the RWI. We focus on   
the co-rotation region 
since this is where relative density perturbations are largest   
and vortex-formation is expected.  

\begin{figure}[!t]
   \centering
   \includegraphics[width = \linewidth,clip=true,trim=0cm 2cm 0cm
     1.0cm]{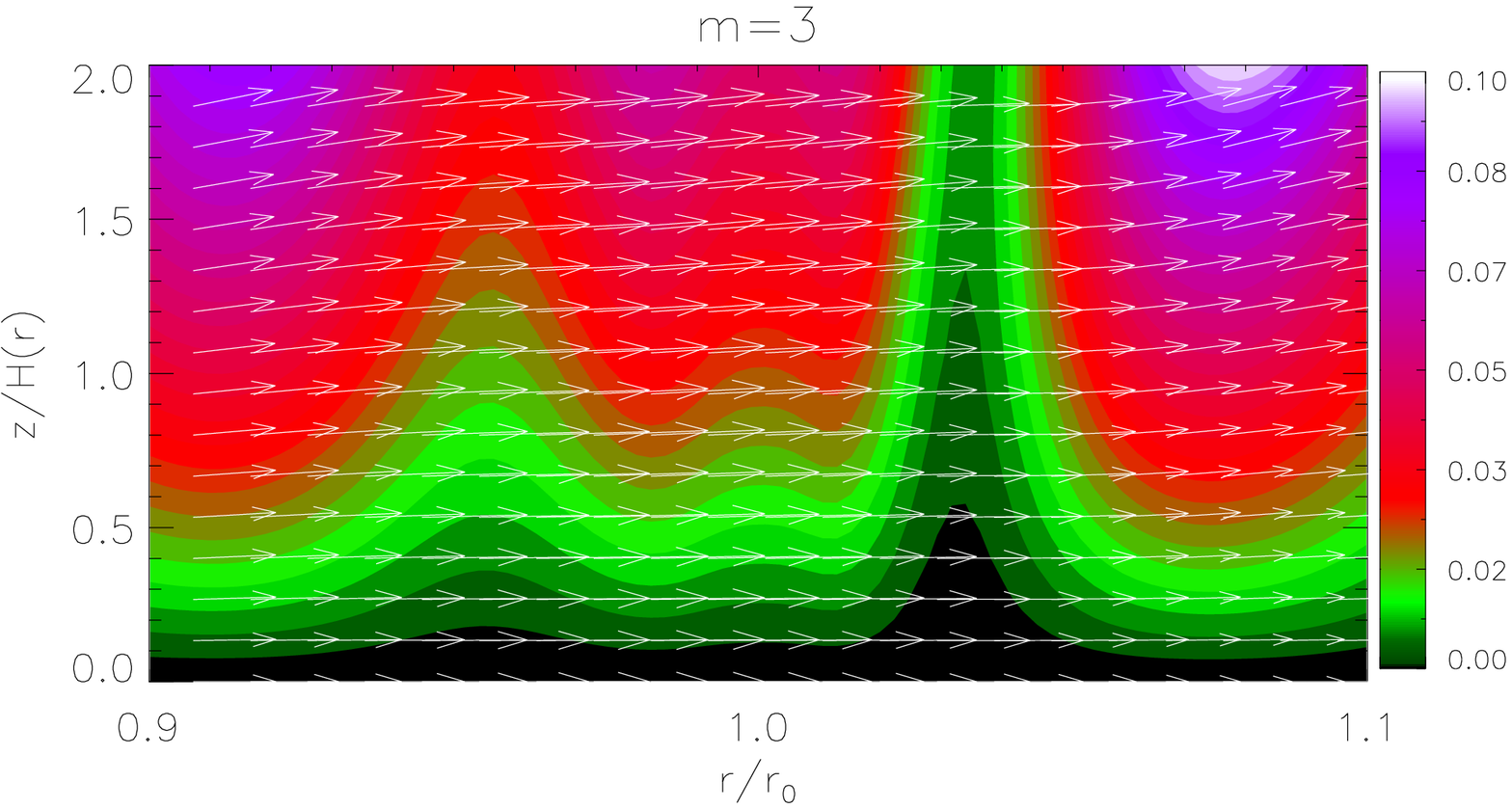} 
   \includegraphics[width = \linewidth,clip=true,trim=0cm 2cm 0cm
     1.0cm]{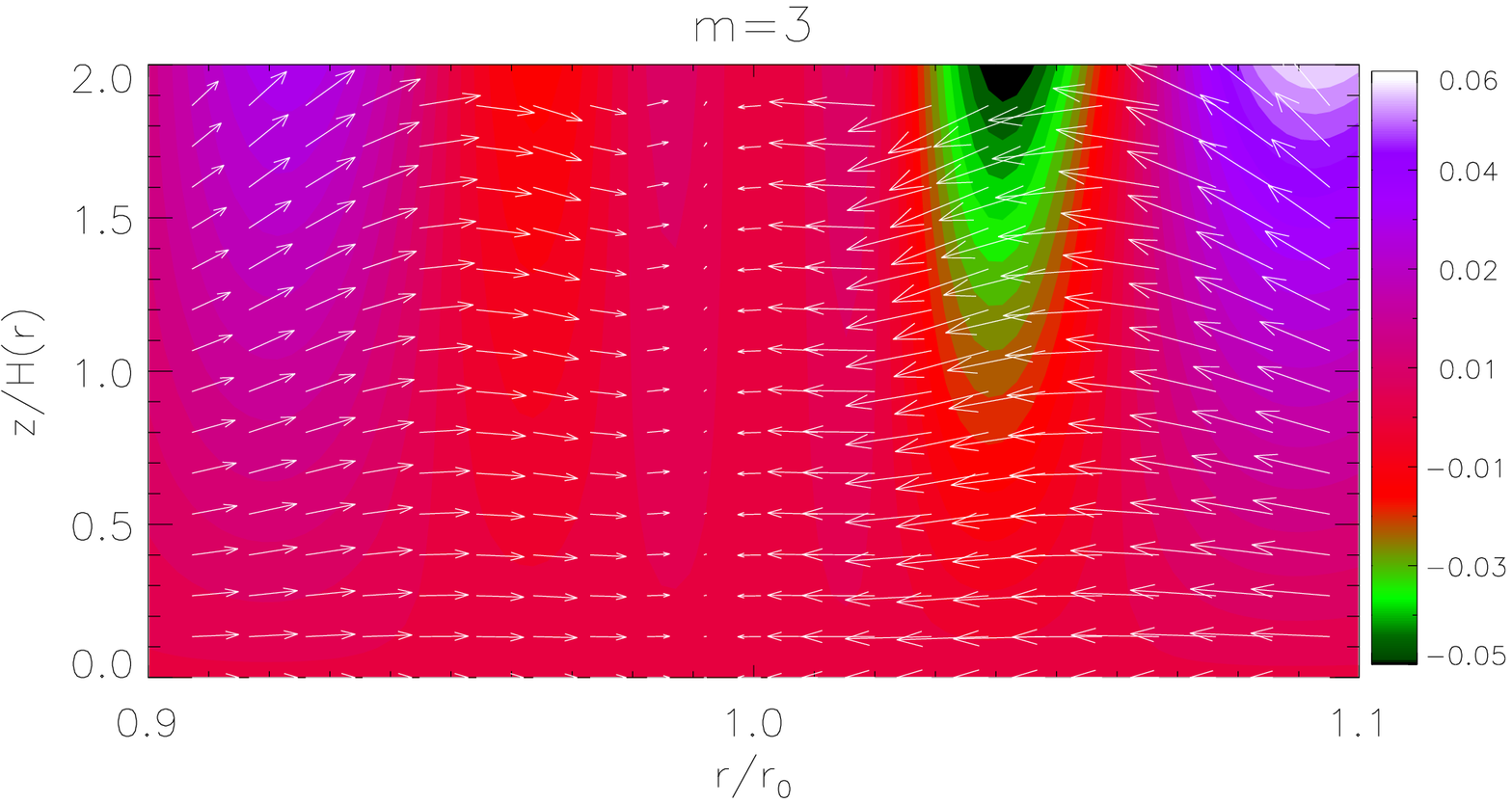} 
   \includegraphics[width = \linewidth,clip=true,trim=0cm 0.5cm 0cm
     1.0cm]{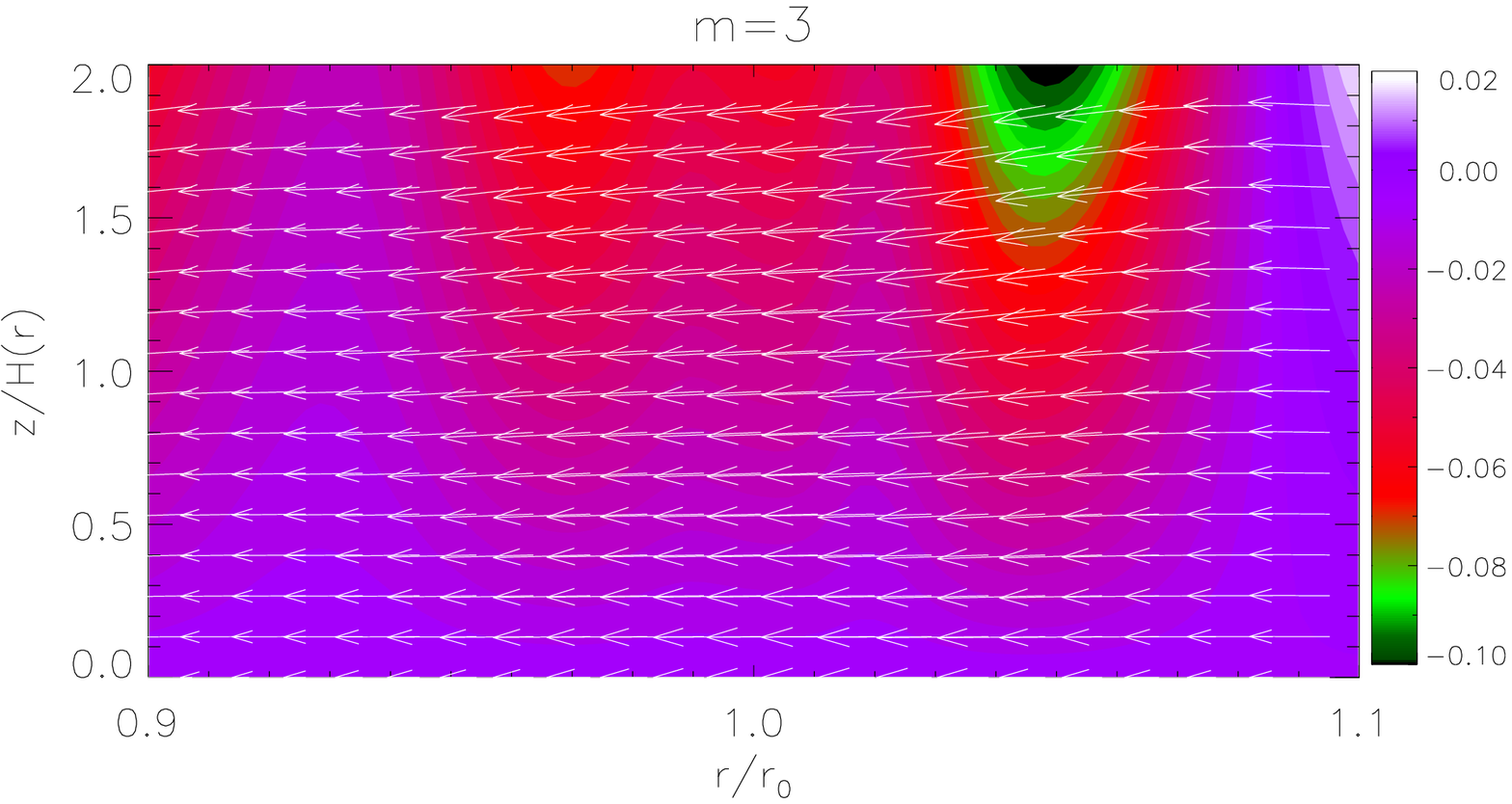} 
   \caption{Vertical velocity field (contours) for the $m=3$ mode in
     the locally isothermal disk with $h=0.07$,  in the $(r,z)$ plane
     at azimuths $\theta\equiv m(\phi-\phi_0) = 0.2\pi$ (top), $0$
     (middle) and $-0.2\pi$ (bottom). Arrows indicate the perturbed velocity field projected 
     onto this plane. 
   \label{vertical_flow_rz}}
 \end{figure}

Fig. \ref{vertical_flow_rz} shows the perturbed vertical velocity
field in the $(r,z)$ plane, at several azimuths. Since the largest
contribution to $\dd v_z$ comes from the $l=2$ term in the expansion
for $W$ (Eq. \ref{hermite_expand}), the magnitude of $\dd v_z$
increases linearly with $z$. 

Ahead and behind the vortex core, the flow just
follows the anti-cyclonic motion, with radial variations in $\dd v_z$
being negligible. At $\phi=\phi_0$ there is also very little vertical 
motion for $z\lesssim0.5H$, but there is upwards motion at 
$r=0.9,\,1.1$, i.e. the edge of the vortex (see
Fig. \ref{2dflow}). This can affect how dust particles are collected 
by RWI vortices.  


\begin{figure}[!t]
  \centering
  \includegraphics[width = \linewidth,clip=true,trim=0cm 0cm 0cm
    1.0cm]{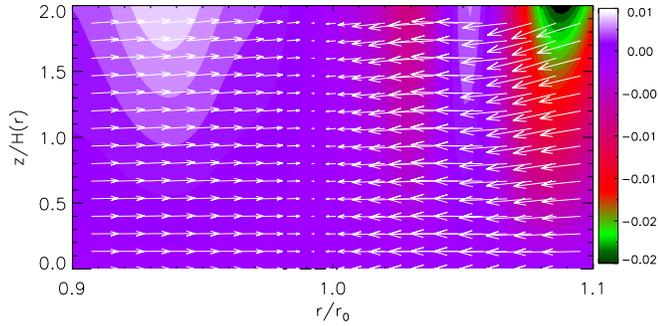} 
  \caption{Vertical velocity field (contours) for the $m=2$ mode 
    in the locally isothermal disk with $h=0.07$. The slice is taken
    at the azimuth $\phi=\phi_0$. This figure is to  be compared with
    the middle plot in Fig. \ref{vertical_flow_rz}.  
    Arrows indicate the perturbed velocity field projected 
    onto this plane. 
    \label{vertical_flow_rz_m24}}
\end{figure}

For comparison, Fig. \ref{vertical_flow_rz_m24}  
shows the vertical flow for the $m=2$ mode. This flow is more
two-dimensional than the fiducial case above. This is expected for
decreasing $m$ \citep[see, e.g.][]{papaloizou85,goldreich86}.  
It also appears 
qualitatively different (e.g. downwards flow at $r=1.1$ instead of
upwards as see for $m=3$). We typically find locally isothermal disks to
display a wider range of flow patterns around co-rotation than
polytropic disks presented later, which show generic patterns.

\begin{figure}[!t]
   \centering
    \includegraphics[width = \linewidth,clip=true,trim=0.0cm 0.5cm
      0.0cm 1.0cm]{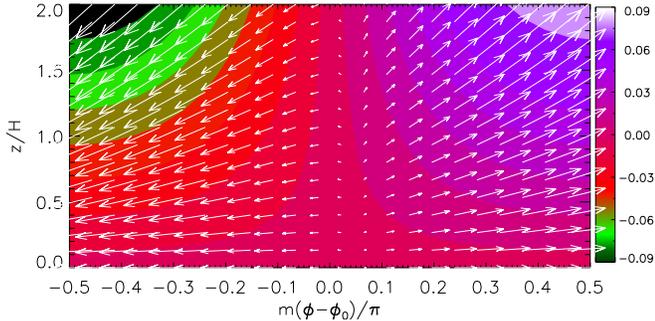} 
   \caption{Vertical velocity field (contours) for the $m=3$ mode in the locally
     isothermal disk with $h=0.07$, in the $(\phi, z)$ plane at the
     radius $r=r_0$. Arrows indicate the perturbed velocity field
     projected onto this plane.  
   \label{vertical_flow_phiz}}
 \end{figure}

Finally, Fig. \ref{vertical_flow_phiz} shows the perturbed vertical
velocity in the $(\phi, z)$ plane at $r=r_0$. Vertical motion is
upwards ahead of an anti-cyclonic vortex and downwards behind it. The
vertical velocity can be comparable to the perturbed azimuthal 
velocity, so the perturbed flow is fully three-dimensional in this plane.  
However, the vortex center $(r_0,\phi_0)$ remains in vertical 
hydrostatic balance. This is not the case for polytropic disks.


\subsection{Dependence of vertical flow on instability 
  strength}\label{iso_3d_effects}  
We now assess how the three-dimensionality of the flow in the
co-rotation region varies with instability strength. We examine 
ratio $\avg{|\dd v_z|}/\avg{|\dd v_r|}$  
, where $\langle\cdot\rangle$ denotes averaging over  
$r\in[0.9,1.1]$ and $z\in[0,2H]$ at fixed azimuth  $\phi=\phi_0$. 
In calculating this ratio, we ignore $W_{l>2}$ because the dominant
contribution  to $\dd v_r$ and $\dd v_z$ comes from $W_0$ and $W_2$ 
respectively. This ratio is large if there is significant vertical
motion.

\begin{figure}[!t]
   \centering
    \includegraphics[width = \linewidth,clip=true,trim=0.0cm 0.0cm 0.0cm 0.0cm]{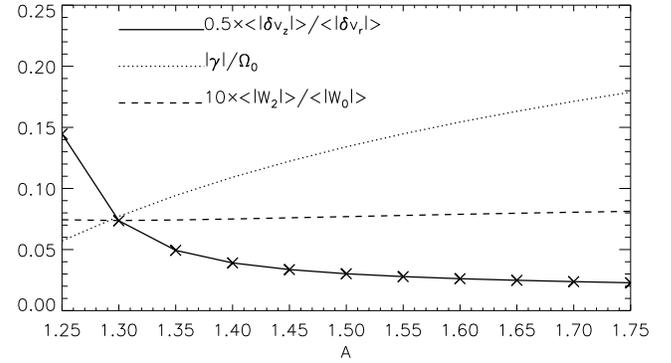}
    \caption{Average magnitude of vertical velocity (solid), in the 
      co-rotation region of the RWI in the locally isothermal
      disk, as a function of bump amplitude $\amp$ at fixed
      aspect-ratio $h$. Also shown are the normalized amplitude of
      $W_2$ in this region (dashed) and the growth rates (dotted). 
      \label{iso_power_amp}}
\end{figure}

Results are shown in Fig. \ref{iso_power_amp}, where
the bump amplitude $\amp$ is increased at fixed $h=0.07$. Growth rates
increase with $\amp$, which is expected from previous works
\citep{li00}, but the flow actually becomes \emph{less}
three-dimensional with increasing instability strength. 


In the co-rotation region where $\sbar\sim \ii\gamma$, we expect 
from the linearized equation of motion that
\begin{align}\label{dvz_corot}
  |\dd v_z| \sim \frac{c_s^2}{H}\left|\frac{W_2}{\gamma}\he_2^\prime\right|.
\end{align}
$|\dd v_z|$ scales with $1/|\gamma|$, so that increasing growth rates 
contributes to decreasing $|\dd v_z|$. Thus, the 
flow in the co-rotation region does not necessarily become more
three-dimensional with increasing $\amp$.  

\begin{figure}[!t]
  \centering
  \includegraphics[width = \linewidth,clip=true,trim=0.0cm 0.5cm
    0.0cm 1.0cm]{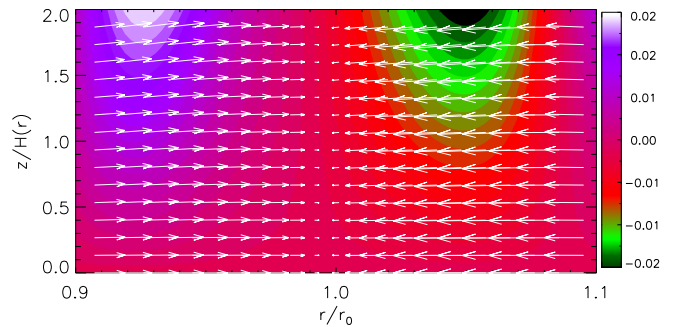} 
  \caption{Same as Fig. \ref{vertical_flow_rz} but  for a disk with
      $\amp=1.6$. The slice is taken at $\phi=\phi_0$. 
    \label{vertical_flow_rz_amp1d6}}
\end{figure}

It is clear from
Fig. \ref{iso_power_amp} that three-dimensionality decreases because
of increasing $|\gamma|$ since $\avg{|W_2|}/\avg{|W_0|}$ varies
weakly. We demonstrate this in
Fig. \ref{vertical_flow_rz_amp1d6},  which shows that in the disk with
$\amp=1.6$ the flow is mainly horizontal. As in the fiducial case with
$\amp=1.25$, there is little motion at $r=r_0$.

Fig. \ref{iso_power_amp}---\ref{vertical_flow_rz_amp1d6} shows that in
the locally isothermal disk, more unstable modes are also  more
two-dimensional (in the co-rotation region). $|W_2|$ remains a small
fraction of $|W_0|$ and $|\dd v_z|$ is largely affected by $|\gamma|$.

However, $|\gamma|$ can be obtained by just solving the 2D
problem. Thus, we could have anticipated the trend of $|\dd v_z|$ in
Fig. \ref{iso_power_amp} based on only 2D
calculations, with the assumption that changes in $|W_2|$ are less
significant than  the increase in $|\gamma|$. 
The above explicit calculation  confirm this, suggesting we  
interpret the RWI as predominantly a 2D instability and that 
three-dimensional effects on the RWI are small (for low $m$). 
We further illustrate these points with polytropic disks below.





\section{Results: polytropic disks}\label{polytropic}
Our fiducial polytropic disk has polytropic index $n=1.5$.  
In the absence of a bump, a surface density profile $\propto r^{-1/2}$
gives a constant aspect-ratio ($H\propto r$). The bump parameters 
are set to $\amp=1.4$ and $h=0.14$. Recall that for polytropic disks,
$H$ is the disk thickness and $h$ is the aspect-ratio at $r_0$. 

\begin{figure}[!t]
   \centering
   \includegraphics[width = 1.0 \linewidth]{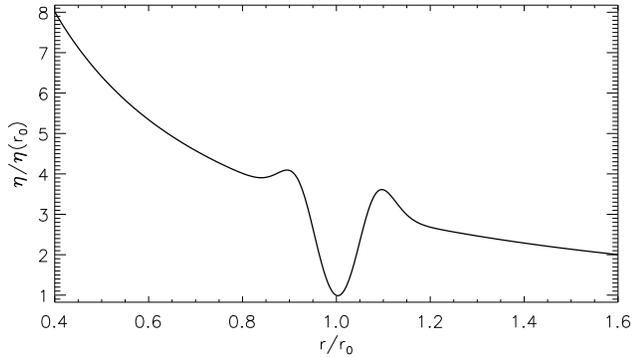}
   \caption{Background profile of the fiducial polytropic disk with
     $n=1.5,\, \amp=1.4$ and $h=0.14$, in terms of the vortensity.   
     \label{basic_poly}}
\end{figure}

Although the surface density enhancement is relatively large, it
corresponds to only $\simeq 9\%$ enhancement of the disk thickness at
$r_0$. The background disk is shown in Fig. \ref{basic_poly} in terms
of the vortensity profile. The fiducial disk has a global
vortensity gradient ($\eta\propto r^{-1}$ away from $r_0$),  but it
is the local minimum that drives instability. The epicycle frequency
is such that  $\mathrm{min}(\kappa^2/\Omega_k^2) = 0.47$.  



\subsection{Solution examples}
Eigenfrequencies for the fiducial case are shown in 
Table \ref{poly_freq}.  The modes of interest are those
with disturbance amplitudes largest near $r_0$, 
which were found to correspond to  $m\leq 4$. 
These modes have effectively 
the same growth rates in 2D and 3D. This  gives confidence that the 
RWI is an 2D instability.   We will consider $m=3$ below in order
  to compare with locally isothermal disks. The $m=3$ growth rate is
  only slightly smaller than the most unstable $m=4$ mode. In 
  co-rotation region, low $m$ modes are also insensitive to radial boundary
  conditions\citep{lin11a}.

\begin{deluxetable}{ccc}
\tablecaption{Eigenfrequencies in the $n=1.5$
  polytropic disk\label{poly_freq}} 
\tablehead{\colhead{$m$} & \colhead{$-\sigma_R/(m\Omega_0)$} & \colhead{$-10^2\gamma/\Omega_0$} \\
\colhead{} & \colhead{} & \colhead{} }
\startdata
1 & 0.9930  (0.9930) & 4.4900  (4.4907) \\

2 & 0.9934 (0.9934) & 8.2793  (8.2867) \\

3 & 0.9941 (0.9941) & 10.769 (10.793) \\

4 & 0.9947  (0.9946) &  11.594 (11.591)\\

5 & 0.9952 (0.9945) & 10.646  (10.861) \\

6 & 0.9954 (0.9950)  & 8.0092  (8.5802)  \\
\enddata
\tablecomments{
  Values in brackets were obtained from the 2D problem.}
\end{deluxetable}

\begin{figure}[!t]
   \centering
   \includegraphics[scale=.425,clip=true,trim=0cm 0.1cm 0cm 0cm]{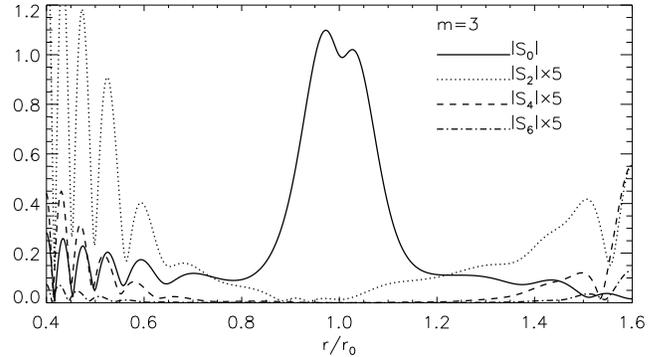}
   \caption{Radial functions $S_l$ for the $n=1.5$ fiducial polytropic disk. 
    These are normalized by $|S_0(r_0)|$. The $l>0$ modes have also
    been magnified in order to compare its radial structure with
    $S_0$.   
     \label{poly_2d_3d_func}}
 \end{figure}

Fig. \ref{poly_2d_3d_func} shows the $S_l$ functions for the $m=3$ 
case. These are similar to the locally isothermal disk
(Fig. \ref{2d_3d_func}).  
We typically find the $l>0$ radial functions to have larger
amplitudes (compared to $l=0$) in the polytropic disk than in locally
isothermal disks. $S_{l>0}$ have small but non-zero 
amplitudes near co-rotation, and their amplitude in the wave-like
regions are at most $\simeq 20\%$ of $|S_0(r_0)|$.     

In the wave-like region, $|S_{l>0}|$ can be comparable or larger than
$|S_0|$. We found the solution in the wave regions more strongly 
affected by boundary conditions than in locally isothermal disks.  

We remark that for $m=5,\,6$, $S_0$ no  
longer has largest disturbance amplitude around $r_0$,  
because radial confinement around co-rotation requires low
$m$ \citep{lin11a}, unless the vortensity minimum is very deep.  
At sufficiently large $m$ (which depends on parameters), the modes are dominated by  the
wave-like region (much like the $m=5$ mode in locally isothermal disks, see
Fig. \ref{2d_3d_func}). Boundary conditions are likely to play a role
here, but they are not the vortex-forming RWI modes of interest.

\subsection{Vertical structure}
We now examine the $m=3$ mode in more detail. 
The flow in the  
$(r,\phi)$ plane is similar to the locally isothermal disk. However,
consistent with the previous section, vertical motion was found to be
more prominent in the polytropic disk. 

\begin{figure}[!t]
   \centering
   \includegraphics[width = \linewidth,clip=true,trim=0cm 0.5cm 0cm
     1.0cm]{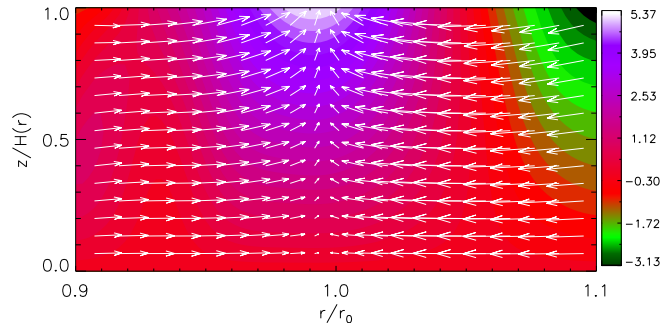} 
   \caption{Vertical velocity field for the $m=3$ mode in the $n=1.5$
     fiducial polytropic disk. The slice is taken in the $(r,z)$ plane
     at the azimuth $\phi=\phi_0$. Arrows are the perturbed velocity
     field projected onto this plane. 
   \label{poly_vertical_flow_rz}}
 \end{figure}

As before we focus on the region $r\in[0.9,1.1]$. 
Fig. \ref{poly_vertical_flow_rz} shows upwards vertical motion  
at the vortex core and is largest near $z=H$. The flow for
$z/H\lesssim0.5$ and/or away from $r_0$ is essentially horizontal. 
The converging flow pattern in Fig. \ref{poly_vertical_flow_rz} is
consistent with $(r_0,\phi_0)$ being an over-density. At
the vortex core, upwards motion makes sense since the midplane is
reflecting. It also implies an increase in disk thickness at
$(r_0,\phi_0)$. 

The background polytropic disk becomes thicker at $r_0$ (i.e. $H$
varies on a local scale). Fluid moving  
into the vortex core finds itself in a region of larger vertical
extent. Upwards motion enhances the disk thickness,
consistent with enhanced pressure and with the RWI vortices being 
over-pressure regions. 

In the locally isothermal disk, it is difficult to directly associate
vertical motion with enhanced pressure as above, since the scale-height is 
prescribed to vary on a global scale  and it remains unperturbed. 
Hence, vertical motion at $(r_0,\phi_0)$ was not
seen in locally isothermal disks.   


We have also  examined the vertical flow in the polytropic disk for other $m$ ($\leq4$),
but found similar flow structure. This is unlike the locally isothermal
disk which can display a range of vertical flow pattern depending on
$m$. (Fig. \ref{vertical_flow_rz}---\ref{vertical_flow_rz_m24}). This
hints that there is a  physical reason why polytropic disks tend to
have positive vertical velocity at $r_0$.  We return to this point
later.

\begin{figure}[!t]
   \centering
    \includegraphics[width = \linewidth,clip=true,trim=0.0cm 0.5cm 0.0cm 1.0cm]{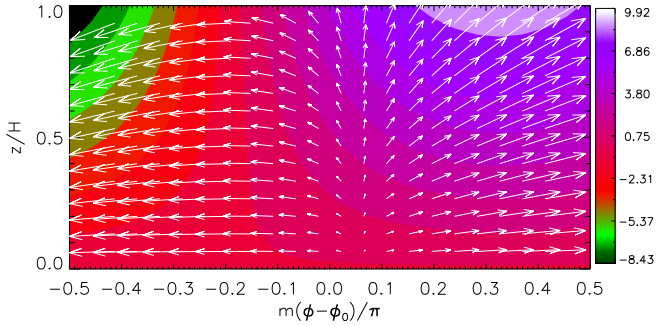}
   \caption{Vertical velocity field for the $m=3$ mode in the $n=1.5$
     fiducial polytropic disk. The slice is taken in the $(\phi,z)$
     plane at radius $r=r_0$. Arrows are the perturbed velocity field
     projected onto this plane. 
   \label{poly_vertical_flow_phiz}}
 \end{figure}

Lastly, Fig. \ref{poly_vertical_flow_phiz} shows the vertical flow in the
$(\phi,z)$ plane at $r=r_0$.  The flow is similar to that in the
locally isothermal disk  (Fig. \ref{vertical_flow_phiz}) except that
the region $\phi\sim\phi_0$ is not in vertical hydrostatic
equilibrium. 

\subsection{Effect of $h$ and $\amp$}\label{poly_3d_effects}
We measure the three-dimensionality of the flow in the co-rotation
region in the same way as in \S\ref{iso_3d_effects}, but here
the averages are taken over the finite vertical extent of the disk.

\begin{figure}[!t]
   \centering
    \includegraphics[width = \linewidth,clip=true,trim=0.0cm 0.0cm 0.0cm 0.0cm]{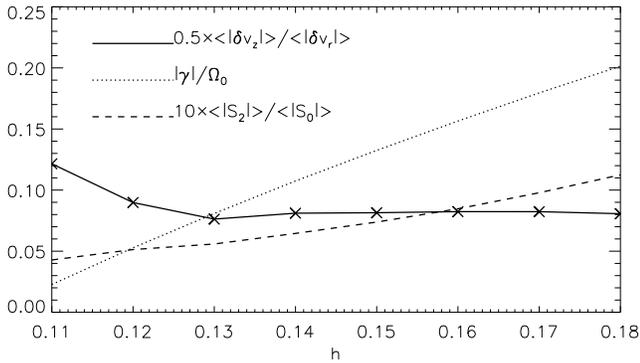} 
   \caption{Effect of $h$ on three-dimensionality of the co-rotation
     flow (solid). 
     Also shown are the normalized amplitude of $S_2$ in this region
     (dashed) and the growth rates (dotted). The increase in growth rates with $h$ is expected 
     because the RWI is driven by pressure forces \citep{li00}.   
   \label{poly_power_h}}
 \end{figure}

\begin{figure}[!t]
   \centering
    \includegraphics[width = \linewidth,clip=true,trim=0.0cm 0.0cm 0.0cm 0.0cm]{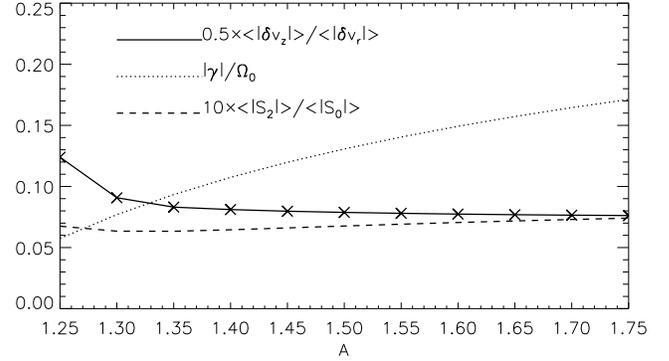}
   \caption{Same as Fig. \ref{poly_power_h}, but as a function of bump amplitude $\amp$. 
   \label{poly_power_amp}}
 \end{figure}

Fig. \ref{poly_power_h}---\ref{poly_power_amp} show results from
calculations with variable $h$ (at fixed $\amp=1.4$) and variable
$\amp$ (at fixed $h=0.14$), respectively. The range of growth rates
are similar to the cases examined for the locally isothermal disk 
 (see Fig. \ref{iso_power_amp}). $|\gamma|$ and $ \avg{|X_2|}$
also behave similarly.  




As in locally isothermal disks,
Fig. \ref{poly_power_h}---\ref{poly_power_amp} shows that the
three-dimensionality of the flow decreases with instability strength,
but less rapidly in polytropic disks. Overall, $\avg{|\dd
  v_z|}/\avg{|\dd v_r|}$ does not vary much, consistent with our
findings that the vertical flow structure, such as 
Fig. \ref{poly_vertical_flow_rz}---
\ref{poly_vertical_flow_phiz}, to be generic. Such plots  are
qualitatively similar across the range of  $h$ and $\amp$ 
considered. The vertical flow at the vortex core is  
always upwards. 

When the spatial average is taken over $r\in[0.98,1.02]$, we found
$\avg{\dd v_z}/\avg{|\dd v_r|}$ maximizes at $h=0.16$ for fixed $\amp$
and at $\amp=1.6$ for fixed $h$.  However, its values are of similar
size: $\avg{\dd v_z}/\avg{|\dd 
  v_r|} \simeq 0.44$---$0.65$   
and $\avg{\dd v_z}/\avg{|\dd v_r|}\simeq 0.53$---$0.65$ 
for variable $h$ and $\amp$, respectively. A reason for such 
insensitivity is that the above calculations have fixed 
polytropic index $n$, thereby fixing the fluid properties. Below, we show
that varying $n$ affects the vertical flow. 


\subsection{Other polytropic indices}\label{varn}
The polytropic index $n$ not only affects 
the magnitude of the bump in the background disk thickness 
but also the compressibility of the fluid. 
An isothermal fluid can be considered a polytrope with 
$n\to\infty$ and is highly compressible, while $n=0$ corresponds to an
incompressible fluid. Thus increasing $n$ also increases
compressibility.

\begin{figure}[!t]
   \centering
    \includegraphics[width = \linewidth,clip=true,trim=0.0cm 0.0cm 0.0cm 0.0cm]{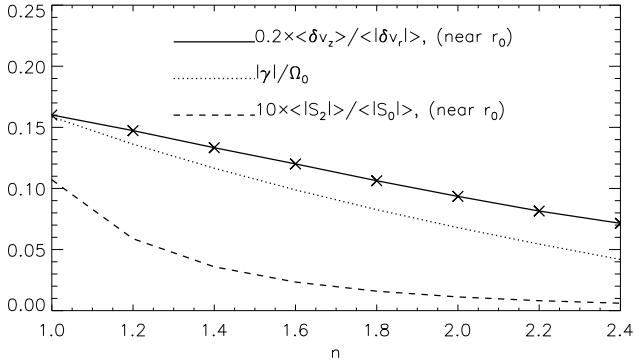}
    \caption{Dependence of the vertical flow at the vortex core  
      on the polytropic index $n$ (solid). The bump amplitude is fixed to $\amp=1.4$ and $h=0.14$. Also shown
      are the growth rates (dotted) and amplitude of $S_2$ (dashed). The mode is $m=3$.   
   \label{poly_power_n}}
\end{figure}

For polytropic disks we identified vertical flow at the vortex core. 
Here, we focus on this region and take radial averages over
$r\in[0.98,1.02]$. Fig. \ref{poly_power_n} show calculations for
$n\in[1,2.4]$. As $n$ decreases, instability strength increases and
the vertical flow at $r_0$ noticeably increases, so the motion becomes
more three-dimensional. This is qualitatively
different from varying $h$ or $\amp$, where the vertical flow
at the vortex core remain of similar size. 

At the co-rotation radius, which is close to $r_0$, 
the vertical velocity is 
\begin{align}
  |\dd v_z| &\sim \left|\frac{S_2}{\gamma H_0}\C_2{^\prime} \right| = \left|(4n^2 - 1)\frac{zS_2}{\gamma H_0^2} \right|.
\end{align}
$H_0$ is constant for fixed $h$. The factor $|(4n^2-1)/\gamma|$
decreases with decreasing $n$, which by itself would reduce the vertical
velocity. Fig. \ref{poly_power_n} shows this is not the
case. The increase in $|S_2|$ with decreasing $n$ overcomes this
effect. 

\begin{figure}[!t]
   \centering
   \includegraphics[width = \linewidth,clip=true,trim=0cm 2cm 0cm
     0.0cm]{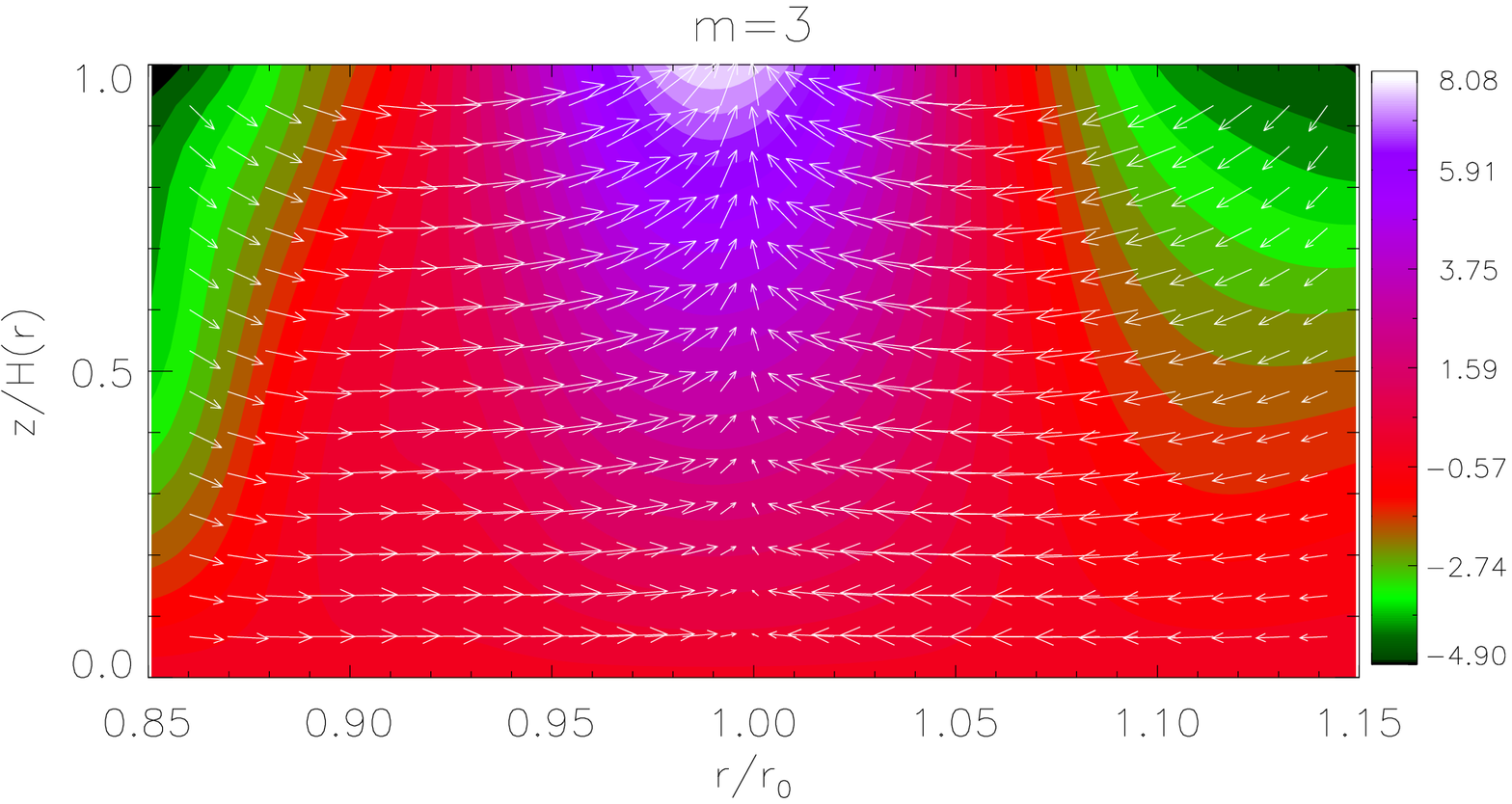}
   \includegraphics[width = \linewidth,clip=true,trim=0cm 0.5cm 0cm
     1.0cm]{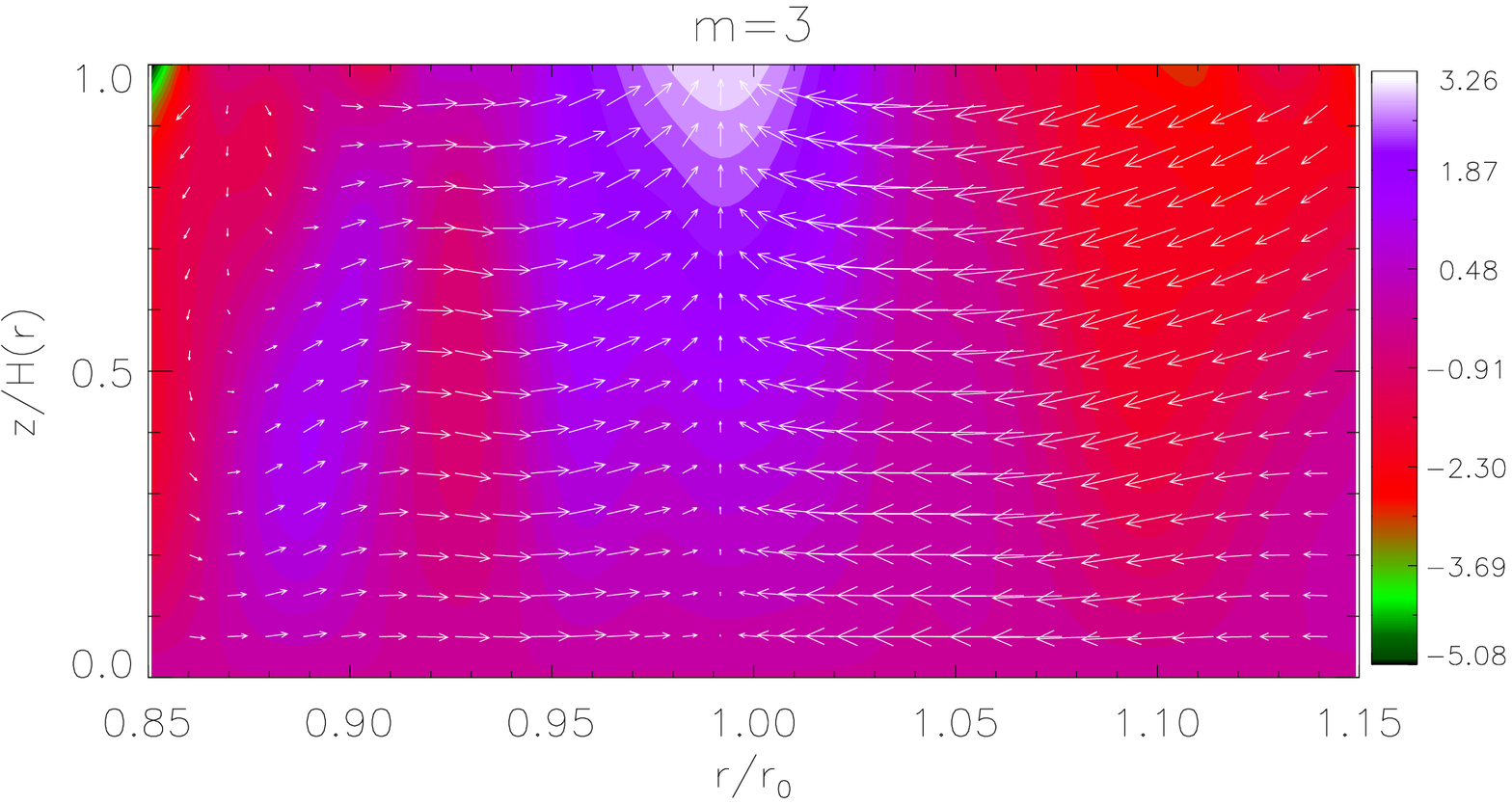}
   \caption{Comparison between vertical velocity (contour) in a
     disk with polytropic index $n=1$ (top) and $n=2$ (bottom). Arrows
     indicate the velocity field projected onto this plane. 
   \label{poly_compare_n_dvz}}
 \end{figure}

In Fig. \ref{poly_compare_n_dvz} we compare the flow in the $(r,z)$
plane between $n=1$ and $n=2$. As previously remarked, the flows share
the same qualitative feature: converging towards $r_0$ with upwards
motion near $r_0$. However, for smaller $n$ (stronger instability), 
upwards motion is concentrated at $r_0$ whereas for larger $n$ 
(weaker instability) there is also upwards motion away from
the vortex core. The latter was also seen for locally isothermal
disks, consistent with larger $n$ being more compressible.    

A larger vertical velocity at $r_0$ with decreasing $n$ 
is consistent with variable compressibility. First note that
$|S_2|\ll|S_0|$ in $r\in[0.9,1.1]$ 
so the perturbed enthalpy, radial and azimuthal
velocities are all dominated by $S_0$, which gives 
converging flow towards the vortex core where there is
enhanced  pressure or density. 
We may then ask what vertical motion at $r_0$ is 
compatible with this 2D perturbed flow, as implied by $S_0$?

At the vortex core $(r_0,\phi_0)$, the linearized continuity
equation is approximately  
\begin{align*}
  \p_t(\dd\rho/\rho) \sim -\nabla\cdot\dd\bm{v} -
  \dd v_z \p_z\ln{\rho},  
\end{align*}
where the $\delta$ quantities are regarded as real. If the fluid is
highly compressible (large $n$), then the density at the vortex core
may increase with vertical motion playing no role. That is, the
divergence term on the RHS dominates over the second
($\nabla\cdot\dd\bm{v}$ itself dominated by horizontal velocities).  

However, if the fluid is made less compressible (decreasing $n$), so that
$\nabla\cdot\dd\bm{v}$ is reduced in magnitude,  then the fluid at
$(r_0,\phi_0)$ should move upwards so that $-\dd v_z\p_z\ln\rho>0$ contributes
to increasing the density. For $n\ll 1$, the fluid becomes
incompressible so that $\nabla\cdot\dd\bm{v}$ is negligible. Then 
the density can only increase by the fluid moving upwards, increasing
the disk thickness and accommodating  more material.

It is important to note that in the above argument, we  
deduced vertical motion by imposing the 2D solution in the
three-dimensional disk. Effectively, we regarded $S_0$ is a source for 
$S_2$, and that $S_2$ has no back-reaction on $S_0$. This
interpretation may not work for general disturbances, however. Here it
is justified by the fact that $|S_2|\ll|S_0|$ from the numerical
calculations. Calculations where the disk is truncated by
setting $r_i=0.7,\,r_o=1.3$, thereby excluding the wave-like
regions in $S_l$, show similar upwards motion. This indicates that $S_0$
induces $S_2$ locally. 




%
%


\section{Disks with $\kappa^2<0$}\label{meheut}
\cite{meheut10} performed the first nonlinear hydrodynamic simulations 
that showed evidence for the RWI in a 3D polytropic disk.  
Their fiducial calculation showed
the development of a $m=1$ anti-cyclonic vortex which survived many
orbits. 

 Indeed, the consideration of polytropic disks in this
  paper was originally inspired by these simulations, but it 
  turns out that the disk model employed by \cite{meheut10} has a
  region where $\kappa^2<0$. Motivated by this feature, in this 
  section we use  \cite{meheut10}'s disk model to explore 
  the 3D RWI when $\kappa_0^2<0$. We find that the RWI can be quite
  different to those described previously (where $\kappa^2>0$ everywhere).


It is straight forward to adapt our setups to models 
used by \citeauthor{meheut10}. They considered a $n=1.5$ polytropic 
disk, occupying $r\in[r_i,r_o]=[1,6]$, and specified the 
midplane density to be a power law ($\rho_0\propto r^{-1/2}$) 
with a Gaussian bump. Their bump in midplane density has the same functional form
as that used for surface density in our models
(Eq. \ref{init_sigma}), so  $\mathcal{A}$ is now the bump amplitude
in midplane density. The bump is located at $r_0=3$ with width
$\Delta r = 0.1r_i$. The calculations presented below employed
$N_r=768$ grid points, on account of the larger disk compared to
previous models.  

 We will consider the $m=1$
  mode below. Calculations were done for 
  $m\leq6$, which gave similar growth rates when 
  $\kappa_0^2<0$, but provided 
  $\mathcal{A}$ is chosen to ensure $\kappa_0^2>0$, then higher 
  $m$ modes become dominant (e.g., with $\mathcal{A}=1.15$,
  $m=5$ had the highest growth rate). The latter is 
  qualitatively consistent with very recent numerical simulations
  \citep[][see also Appendix \ref{improved}]{meheut12b}.     

  When $\kappa_0^2>0$, we find similar flow
  structure to that described previously. Having applied the linear
  calculations to a different disk model and recovering similar 
  results gives us confidence in the robustness of the RWI to develop 
  3D.

\subsection{$m=1$ modes} 
In their fiducial setup, \citeauthor{meheut10} adopted a bump amplitude
of $\amp=1.4$. This results in $\kappa^2=0$ at
$r\simeq0.99r_0,\,1.01r_0$ and $\kappa^2_0<0$  
(which is also reflected in their Fig. 9). The disk is therefore
unstable to local axisymmetric perturbations \citep{chandrasekhar61}.

Interestingly, for $\amp=1.4$ we found a $m=1$ mode with 
large  growth rate,  $|\gamma|=0.36\Omega_0$, almost twice
the largest growth rates found previously. Below, we
examine this solution along with a case with $\amp=1.3$, which has  
$\kappa^2>0$ everywhere and growth rate
$|\gamma|=0.05\Omega_0$\footnote{This is comparable to the nonlinear
  simulation.}.  

Despite $\amp$ being similar, the $m=1$ growth rate
for $\amp=1.3$ is much smaller than   
that for $\amp=1.4$. For $\amp=1.3$ we did not find other $m=1$ modes  
with growth rates similar to the $m=1$ mode in $\amp=1.4$.  
Furthermore, for $\amp=1.4$ the quantity $D=\kappa^2 -
\sbar^2$ almost vanishes near $r_0$: 
\begin{align*}
  \mathrm{min}(|D|/\Omega_k^2) = 4\times10^{-3},
\end{align*}
which occurs at $r=1.002r_0$. For $\amp=1.3$, the value above is 
$0.14$. 

\begin{figure}[!t]
   \centering
   \includegraphics[scale=.425,clip=true,trim=0cm 1.8cm 0cm
     0cm]{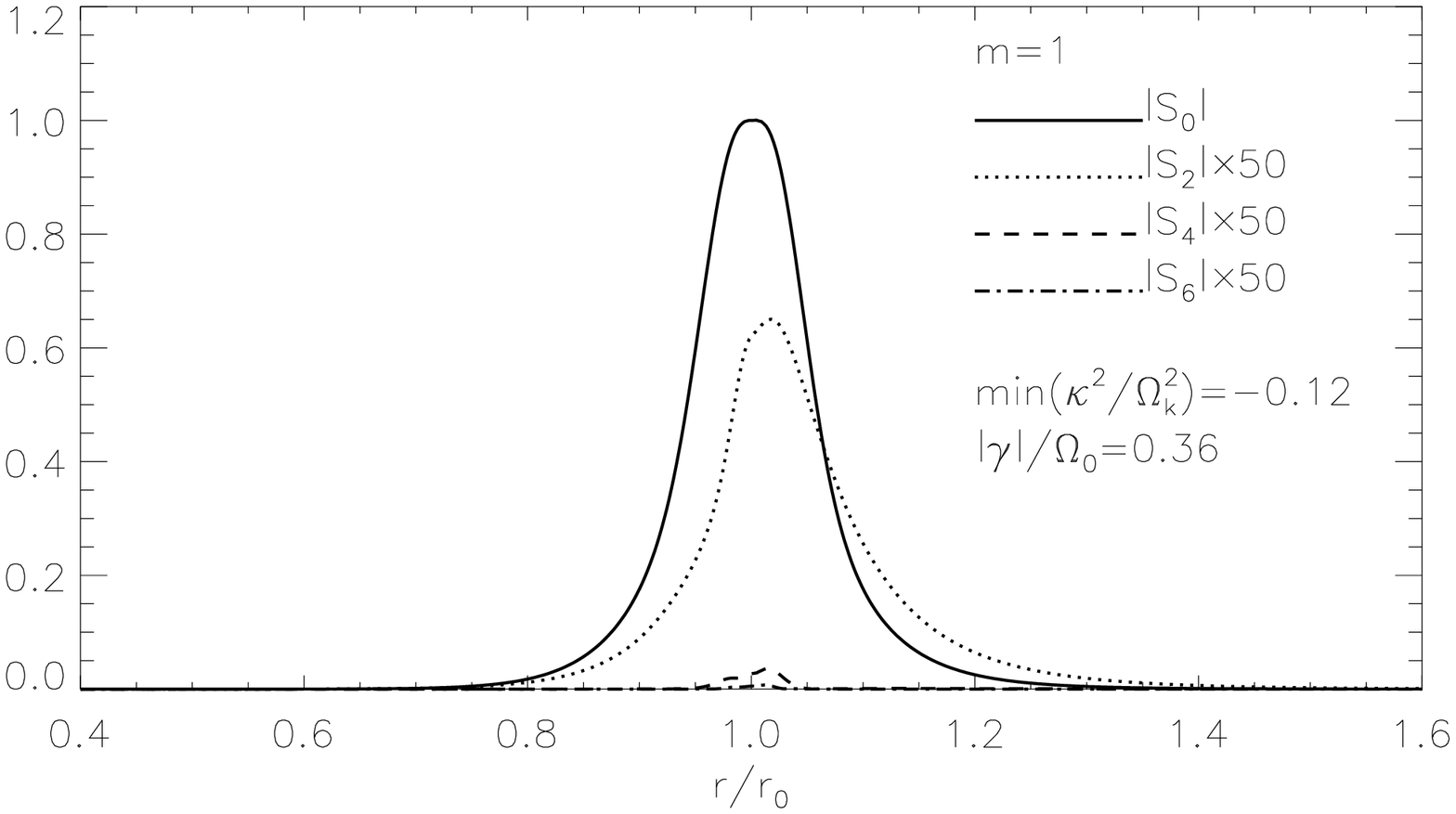} 
   \includegraphics[scale=.425,clip=true,trim=0cm 0.0cm 0cm
     0.2cm]{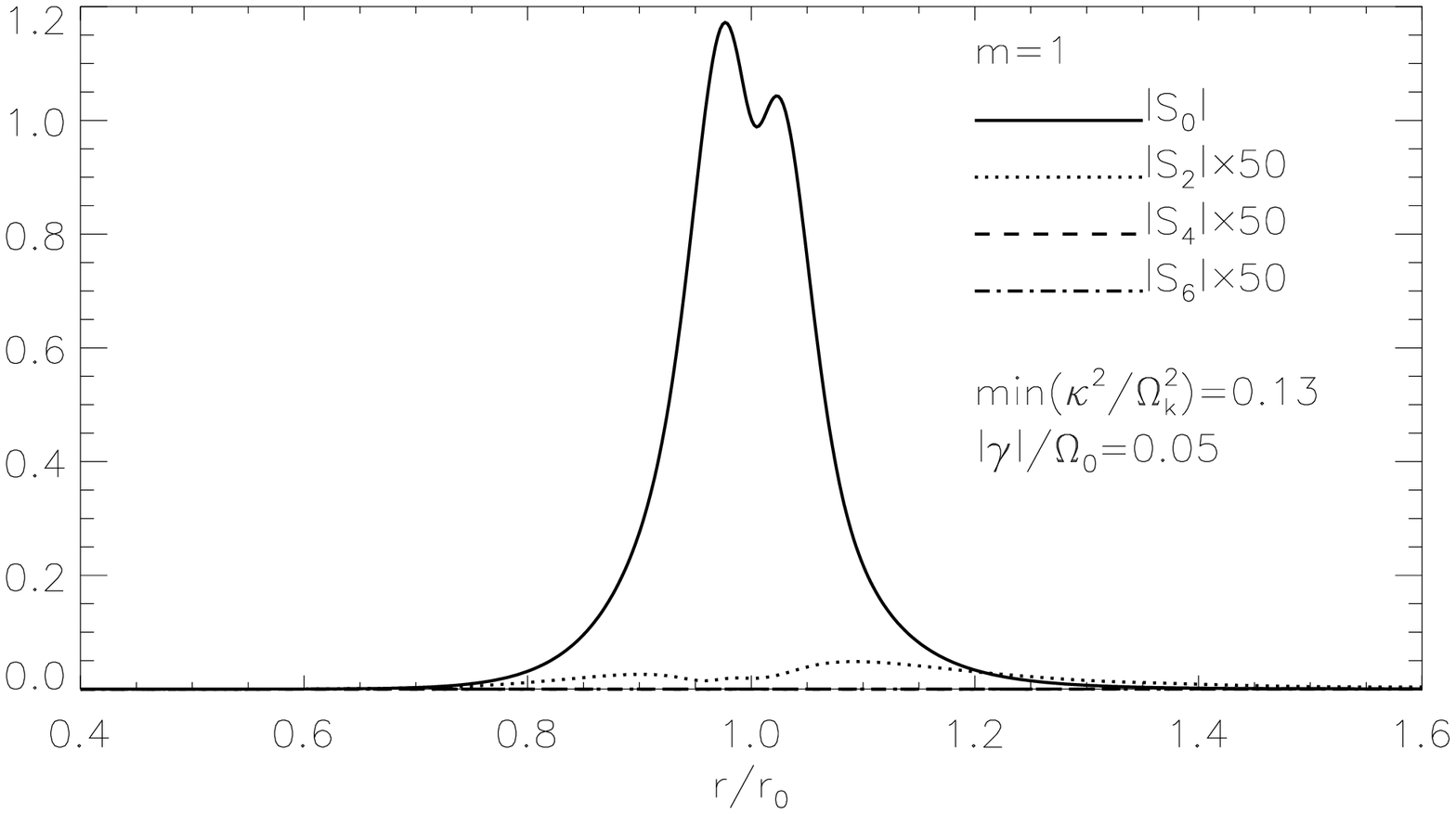} 
   \caption{Radial functions $S_l$ for the polytropic disk model 
     with a bump in midplane density of amplitude
     $\amp=1.4$ (top, \citeauthor{meheut10}'s fiducial setup) and with
     a bump amplitude of $\amp=1.3$ (bottom). For $\amp=1.4$,
     $\kappa_0^2<0$. 
     \label{meheut_func}}
\end{figure}

Fig. \ref{meheut_func} compares the $S_l$ functions for the cases above. 
While the double-peak in $S_0$ for $\amp=1.3$ was also found in
previous sections and also by \cite{li00}, it is absent in $\amp=1.4$.  
The dominant 3D mode is $S_2$, but it is significantly larger in  
$\amp=1.4$ than in $\amp=1.3$. This indicates the vertical flow will also
be qualitatively different.  
 
\begin{figure}[!t]
   \centering
    \includegraphics[width = \linewidth,clip=true,trim=0.0cm 0.5cm 0.0cm 1.0cm]{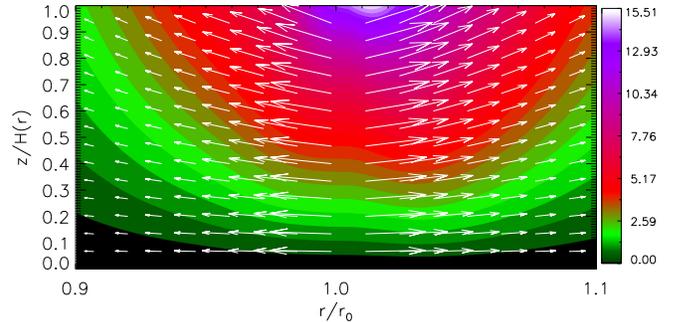}
   \caption{Vertical velocity field for a $m=1$ mode in \cite{meheut10}'s disk model with
    $\amp=1.4$, which results in in $\kappa^2<0$ at $r_0$ and $ D \to
     0$ there as well. The arrows are the perturbed velocity field
     projected onto this plane.  This result is qualitatively different to modes
     with $\kappa^2>0$ (see figures in \S\ref{polytropic}).   
   \label{meheut_vertical_flow_dvz}}
\end{figure}

Fig. \ref{meheut_vertical_flow_dvz} shows the flow pattern at $\phi=\phi_0$ for 
$\amp=1.4$. This result is very different from that for $\amp=1.3$,
which share the same features as our previous setups with $\kappa^2>0$
(e.g. Fig. \ref{poly_vertical_flow_rz}). Note that while
the $S_l$ behave smoothly across $r_0$ (Fig. \ref{meheut_func}),
numerical evaluation of $\dd v_r$ involves a division by $D$, which
is very small near $r_0$ for $\amp=1.4$. Thus, horizontal
velocities may be subject to   numerical artifacts at $r_0$. Despite
this, the direction of radial flow, being inwards for $r<r_0$ and
outwards for $r>r_0$ with a sharp transition at $r_0$, was also found
in \citet[their Fig. 11]{meheut10}.

 Neither $\amp$ produced vertical flow consistent with 
that in \cite{meheut10} where strong downwards flow at
$r_0$ were identified with rolls excited on either side. 
 By contrast the linear solutions have upwards motion
and there is no vortical motion in the $(r,z)$ plane.

Despite using the same disk models, several factors may have 
contributed to the discrepancy between the linear calculation above and
\citeauthor{meheut10}'s simulation.   
These include the treatment of the vertical domain,
  nonlinearities (H. Meheut, private communication) and interaction
  with other $m$ modes in the simulation which cannot be treated in 
  linear theory. 
  
  



There may also be numerical issues in our linear calculation because
of $\kappa^2\leq0$. The RWI is associated 
with the term $\propto(1/\sbar)\p_r(\rho_0\Omega/D)$ and its disturbance is localized about $r_0$. 
This term is $\propto 1/\kappa^2$, which diverges when $\kappa^2\to0$ near $r_0$. Also because
$\kappa^2_0<0$, it allows $D\to0$ at co-rotation as well. We have
performed calculations with lower spatial resolution, so that
numerically $\kappa^2$ and $D$ have larger deviations from zero, but
we found similar eigenfrequencies and flow patterns to the case shown
above. We will further comment on RWI modes with $\kappa_0^2<0$  in 
\S\ref{neg_ksq}.

%

\section{Summary and discussion}\label{summary}
In this paper, we have examined the linear stability of 
three-dimensional, vertically stratified and radially structured
disks. Our calculations are 3D analogs of those presented
by \cite{li00}, in which the Rossby wave instability was studied in 
razor-thin disks. 
 In order to simplify the problem, we assumed  
  the perturbed hydrodynamic quantities have vertical dependence
  that can be decomposed into Hermite or Gegenbauer polynomials. Our
  conclusions therefore apply to such perturbations only.

Our numerical calculations confirm the RWI persists in 3D. For ease of
discussion below, we denote  the full linear solution schematically as 
\begin{align*}
X=Y(r) + \Delta Y(r,z), 
\end{align*}
where $Y$ is the $z$-independent part of the solution and $\Delta Y$ 
is the part that also depends on $z$.  

\subsection{Validity of 2D}
We find the RWI growth rate $|\gamma|$ can be accurately determined 
from the 2D problem alone. 
In other words, instability is associated  
with $Y(r)$. In the region of interest --- the vortensity minimum ---
where vortex-formation is expected, we find $|\Delta Y|\ll|Y|$ so that
enthalpy, radial velocity and azimuthal velocity perturbations 
have essentially no $z$-dependence.  

In fact, weak $z$-dependence 
is expected from earlier studies of accretion tori. 
For slender tori, \cite{papaloizou85} demonstrated the
existence of low $m$ unstable  modes with weak 
$z$-dependence. \cite{goldreich86} also justified the use of 
height-averaged equations for calculating modes in narrow tori, for
which vertical hydrostatic equilibrium was assumed. Although we
considered radially extended disks, their results should  
apply here because the low $m$ RWI modes, relevant to  
vortex-formation, have largest disturbance associated with a narrow 
region about the density bump.   
More recently, \cite{umurhan10} reproduced the RWI in approximate
three-dimensional disk models, in which horizontal velocities have no
vertical dependence. Our numerical results are therefore supported by 
analytic studies above.  

The 2D solution, $Y$, imply anti-cyclonic motion associated with
over-densities, thus we expect the RWI will lead to columnar vortices
in 3D.  The survival of vortices in 3D is then an important issue 
because they may be subject to instabilities \citep{lesur09,lesur10}. 
On the other hand, if there is a continuous source of vortensity 
extremum, such as disk-planet interaction, then vortex-formation
via the RWI could be maintained.

%
%

\subsection{Vertical motion}

Although $|\Delta Y|$ is small in the co-rotation region, it is
nevertheless non-zero. This implies vertical motion growing on
dynamical time-scales, making the flow in the co-rotation
region three-dimensional. We found the nature of the vertical flow  
is affected by the equation of state. 

In polytropic disks the vortex core $(r_0,\phi_0)$ always  
involve upwards motion. For fixed polytropic index $n$, there is
limited variation in the magnitude of vertical flow with respect to
instability strength. However, if the fluid is made less compressible
by lowering $n$, then vertical motion at the vortex core increases.

This result motivates us to interpret vertical motion around
co-rotation as a perturbation to the 2D solution  
\citep{goldreich86}. Recall that $Y$ is the solution to the vertically   
integrated system. It signifies non-axisymmetric enhancements in
surface density at the bump radius. This characteristic feature is
unchanged by the addition of $\Delta Y$ to the 2D solution. We then 
ask how should the disk respond in the vertical direction.    
 
The polytropic disk thickness is directly related to the
surface density (Eq. \ref{poly_thick}). Enhancement of the surface
density therefore imply enhancement in disk thickness, so fluid at
$(r_0,\phi_0)$ moves upwards. If we look in the $(\phi,z)$ plane at
$r_0$, the disk thickness becomes non-axisymmetric. This has already
been observed in nonlinear simulations \citep{meheut11a,meheut11b}. In
these newer simulations, the authors indeed find upwards motion in 
anti-cyclonic vortices. 

Note that the polytropic disk thickness becomes less sensitive
to surface density as $n$ is increased (Eq. \ref{poly_thick}).  
For $n\to\infty$ the disk thickness is independent of surface density 
and there is no need for fluid to move vertically in order to achieve a 
surface density increase. In this case there is no preference for  
vertical velocity of a particular sign at $(r_0,\phi_0)$.   Since
  the fluid behaves isothermally
  as $n\to\infty$, the above is
  consistent with our observation 
  that locally isothermal disks have little vertical motion right at
  the vortex core. In Appendix \ref{consistency} we consider a
  polytropic disk calculation with $n=8$ to check for consistency.

\subsection{RWI with $\kappa^2<0$}\label{neg_ksq}
 We briefly examined the linear 3D RWI 
  in disk where $\kappa^2$ becomes negative at the 
  density bump. This was inspired by the 3D RWI simulations
  presented in \cite{meheut10}, where the disk model 
  had $\kappa_0^2<0$. In this setup we found a $m=1$ linear mode with
  large  growth rate and qualitatively different to modes in disks with
  $\kappa^2>0$ everywhere. 
  In neither case did we reproduce the vertical flow seen in
  \cite{meheut10}, namely downwards flow at the vortex center. 

Most discussions of non-axisymmetric disk instabilities have
assumed $\kappa^2>0$ everywhere, including \cite{lovelace99}'s
original description of the RWI, so that     
Rayleigh's criterion for stability against local axisymmetric
perturbations is satisfied. 

The RWI has been shown to exist for $\kappa_0^2<0$ but its
properties appear different to those in disks with $\kappa_0^2>0$. For 
example, \cite{li00}'s linear calculations indicate a non-smooth
change in growth rate as $\kappa_0^2$ becomes negative (the `HGB' case
in their Fig. 9).  In nonlinear 2D simulations by \cite{li01}, the RWI
also evolves differently depending on whether the growth rate is low
($|\gamma|\sim 0.1\Omega_0$ and $\kappa_0^2 >0$) or high
($|\gamma|\sim0.3\Omega_0$ and $\kappa_0^2<0$). Note the latter 
case has  $|\gamma|$ close to that found in our calculation. We
therefore expect the RWI to differ in 3D depending on
$\sgn(\kappa_0^2)$. This is apparent by comparing our results with 
$\kappa_0^2>0$ to those with $\kappa_0^2<0$.   


 Thus, while \cite{meheut10} is the first 
  demonstration of the 3D RWI, it should be kept in mind that the disk
  model has $\kappa_0^2<0$. An understanding of such modes in 3D
is of theoretical interest, but it is unclear whether or not
protoplanetary disks  will develop sufficiently large pressure
gradients to render $\kappa^2<0$ \citep{yang10}.

\subsection{Outstanding issues}\label{caveats}
The main goal of our study is to demonstrate the linear RWI in 3D and
to identify the nature of associated three-dimensional flow 
structure around co-rotation. However, our study is subject to several
caveats which should be clarified in future work.   
 
\subsubsection{Baroclinic effects}\label{baroclinic}

One issue is that our locally isothermal basic states are not  
in true equilibrium, because we  approximated the rotation profile
to be $z$-independent (Eq. \ref{unperturbed_vphi}). Initializing a
full hydrodynamic simulation this way might boost radial velocities
because of the inexact radial momentum balance. In order for the
angular velocity to be strictly independent of $z$, we must set 
$H\propto r^{3/2}$, which is the globally isothermal disk already
considered by \cite{meheut12}. We do not expect this to make a 
difference from our disks with $H\propto r$, because the
RWI is driven by local variations in disk structure and its 
disturbance is radially confined.  We check this in Appendix
  \ref{consistency}.

Another justification is that for a thin, smooth disk
($\amp=1$) with $H=hr$, the angular velocity is  
\begin{align}
\Omega(r,z) = \Omega_k\left[1 - \frac{h^2}{2}\left(\alpha + 2 +
  \frac{z^2}{2H^2}\right)\right] 
\end{align}
\citep{tanaka02}. 
The difference in angular speed between the gas at the midplane and
gas at $z$ is then  
\begin{align}\label{Delta_Omega}
\Delta\Omega \equiv |\Omega(r,z)-\Omega(r,0)| =
h^2\left(\frac{z^2}{4H^2}\right)\Omega_k 
\end{align}
(a radial density bump does not contribute to this difference). 
Since the gas is contained within a few scale-heights, we have   
$\Delta\Omega/\Omega_k= O(h^2)$. Because $h\ll 1$,  
vertical shear should be unimportant if the dynamics of interest
operate on much faster time-scales, as can be the  case for the RWI
with growth rates $\sim h\Omega_k$. That is, the vortical perturbation
grows much faster than it is sheared apart by
$\Delta\Omega$. We have begun preliminary nonlinear simulations which
confirms vortex formation via the RWI in a locally isothermal 3D disk with constant
aspect-ratio (Lin 2012, in preparation).    

\cite{knobloch86} have pointed out the possibility
of baroclinic instability in the case of $\Omega=\Omega(r,z)$, when there
are radial variations in temperature on the scale of local
scale-heights. This condition is not met in our locally isothermal
disk models because the sound-speed varies on a global scale. 
In more realistic disk models, one might expect that a density bump 
also involves local temperature variations. Baroclinic effects may 
then become important. On the other hand, the RWI may also be enhanced 
because of local temperature gradients \citep{li00}. 
Having  $\Omega=\Omega(r,z)$ means solving the linearized
equations as a PDE eigenvalue problem, which is not simple.

\subsubsection{Boundary effects}

We have restricted our attention to the co-rotation region because this
is where vortex-formation eventually takes place. Distant radial
boundaries do not affect the dynamics in this region significantly (as
checked numerically). However, it is
clear that far away from co-rotation, three-dimensional effects become
increasingly important. This is seen in the polytropic disk as
$|\Delta Y|\sim|Y|$ towards the disk boundaries 
(Fig. \ref{poly_2d_3d_func}). Disturbances associated with the RWI are
therefore three-dimensional beyond the Lindblad
resonances. In order to study these regions, more physically realistic
radial boundary conditions are needed.

Around co-rotation the RWI is a 
global disturbance in $z$, so the upper disk boundary conditions could be
important. The use of orthogonal polynomials means we simply impose a
regularity condition at the upper disk boundary (\S\ref{numerics}). 
This method of solution does not allow us to explore the effect of
other vertical boundary conditions. Again, such a study involves a 
PDE eigenvalue problem, but can reveal to what extent the dominance 
of the 2D solutions found here are influenced by the specific
decompositions employed. This will be the subject of a follow up
paper.   

Nevertheless, we can make some speculations based on results here. 
The vanishing density at the polytropic disk surface 
is likely to provide a reflective upper boundary. This effect may be
important. It might reduce the growth of the RWI if it remains
predominantly a 2D disturbance, because the 2D solution alters the
surface density,  which is directly related to the disk thickness
for a polytrope, but the disk thickness cannot change.   




\acknowledgements

I thank H. Meheut for useful discussions and clarification of their simulation results. 
I also thank O. Umurhan for comments on the first version of this paper.

\appendix

\section{Explicit expressions for the linear operators}\label{expressions}
\subsection{Locally isothermal disks with constant aspect-ratio}
For locally isothermal disks with $H=hr$ with $h$ being a constant and $\Omega$ taken to be a function
of radius only, the
operators governing the linear problem are given by 
\begin{align}
  A_l = & \left[ \frac{2mr\Omega}{\sbar}\frac{d}{dr}\ln{\left(\frac{c_s^2\Sigma\Omega}{D}\right)} 
    -\left(m^2 + \frac{r^2D}{c_s^2}\right) + \frac{lD}{\sbar^2h^2} - lr \frac{d}{dr}\ln{\left(\frac{c_s^2\Sigma}{D}\right)}
    -l(2l-1) + \frac{4m\Omega l}{\sbar}\right] \notag \\
  &+ r^2\frac{d}{dr}\left[\ln{\left(\frac{rc_s^2\Sigma}{D}\right)}\right]\frac{d}{dr} + r^2\frac{d^2}{dr^2},\label{3d_hermite1}\\ 
B_l = & -\left[(l-2) - \frac{2m\Omega}{\sbar} \right] + r\frac{d}{dr},\label{3d_hermite2}\\
C_l = &
   -(l+1)(l+2)\left[ r\frac{d}{dr}\ln{\left(\frac{c_s^2\Sigma}{D}\right)} + l - \frac{2m\Omega}{\sbar}\right]
  -r(l+1)(l+2) \frac{d}{dr}. \label{3d_hermite3}
\end{align}
We have expressed the operators in terms of surface density $\Sigma$
so the above may be  seen to be equivalent to Eq. 21 in
\cite{tanaka02} when their parameter $\mu=d\ln{H}/d\ln{r}$ is set to
unity. 

 These equations are
  approximate because we ignored terms proportional to $\p_z\Omega $ in the 
  governing PDE from which Eq. \ref{3d_hermite1}---\ref{3d_hermite3} 
  are derived. These terms are non-vanishing for exact
  equilibrium if the sound-speed varies with radius, but
  for a thin disk $\p_z\Omega\propto  h^2\ll1$ so we expect them to be
  small. It is worth neglecting them in favor of the 
  one-dimensional operators above, which are much simpler. 
  \cite{tanaka02} gives a more general equation for the linear
  problem which includes $\p_z\Omega$. Their Eq. 11 shows that
  $\p_z\Omega$ contributes to the coefficient of $\p_zW$ as   
  \begin{align}
    \frac{\p W}{\p z}\left[\frac{z}{H^2} +
      \frac{m}{\sbar}\frac{\p\Omega}{\p z} \right] = \frac{z}{H^2}\frac{\p W}{\p z}\left[ 1 -  
      \frac{mh^2q\Omega_k}{\sbar}\right], 
  \end{align}
  where $\p_z\Omega$ is evaluated using \citeauthor{tanaka02}'s Eq. 4 and 
  $q\equiv -d\ln{c_s}/d\ln{r}=0.5$ for disks with constant aspect-ratio (equivalent to Eq. \ref{Delta_Omega} in
  \S\ref{baroclinic}). Near co-rotation the magnitude of the second to first term is  
  \begin{align}
    \left|mh^2q\frac{\Omega_k}{\sbar}\right|\sim
    \frac{mh^2q}{|\gamma/\Omega_k|}. 
  \end{align}
  For the fiducial case in \S\ref{isothermal}, $m=3,\,h=0.07$ and $|\gamma/\Omega_k|\simeq
  0.057$, this ratio is 0.13. We typically find 
  $|\gamma/\Omega_k|=O(h)$, so the second term is a
  factor $mhq\ll1$ smaller than the first for low $m$
  modes. Neglecting it (to arrive at
  Eq. \ref{3d_hermite1}---\ref{3d_hermite3}) is a self-consistent
  treatment.

\subsection{Polytropic disks}

For polytropic disks, we find it most convenient to express the linear operators as 
\begin{align} 
A_0 =& -\wop{1} + \frac{(2\lambda+1)}{2(\lambda + 1)}
\left[ \wop{2} - \wop{4} + \wop{6} - \wop{7} + \wop{8}\right],\\
A_{l>0} =& -\wop{1} + \frac{1}{2(l+\lambda + 1)(l+\lambda -1)}
\left\{ \left( l^2 + 2\lambda l + 2\lambda^2 - \lambda - 1\right)
        \left[\wop{2} + \wop{6} + \wop{8}\right]\right.\notag \\
&\phantom{-\wop{1} + \frac{1}{2(l+\lambda + 1)(l+\lambda -1)h}}
 +l(l+2\lambda)\left[ \lambda\wop{3} - \left(l^2 + 2\lambda l -1\right)\wop{5}\right] \notag\\
&\phantom{-\wop{1} + \frac{1}{2(l+\lambda + 1)(l+\lambda -1)h}}
   -\left[(\lambda+1)l^2 + 2\lambda(\lambda+1)l + 2\lambda^2 - \lambda -1\right]\wop{4}\notag\\
&\phantom{-\wop{1} + \frac{1}{2(l+\lambda + 1)(l+\lambda -1)h}}
   \left.    -(l^2 + 2\lambda l + \lambda -1)(2\lambda + 1)\wop{7} \right\}
-l(l + 2\lambda)\wop{9},\\
B_l =& -\frac{l(l-1)}{4(l+\lambda - 2)(l+\lambda -1)}
\left\{\wop{2} + (l-2)\wop{3} + (l+2\lambda -1)\left[\wop{4} + (l-2)\wop{5}\right]
       \right. \notag\\
&\phantom{-\frac{l(l-1)}{4(l+\lambda - 2)(l+\lambda -1)}h}  
\left. + \wop{6} + (2\lambda + 1)\wop{7} + \wop{8}\right\},\\
C_l =& \frac{(l+2\lambda + 1)(l+ 2\lambda)}{4(l+\lambda + 1)(l+\lambda + 2)}
 \left\{-\wop{2} + (l+2\lambda + 2)\wop{3} 
+ (l+1)\left[\wop{4} - (l+2\lambda + 2)\wop{5} \right]\right.\notag\\
&\phantom{\frac{(l+2\lambda + 1)(l+ 2\lambda)}{4(l+\lambda + 1)(l+\lambda + 2)}h}
\left. - \wop{6} - (2\lambda+1)\wop{7} - \wop{8}
\right\},
\end{align}
where
\begin{align}
\wop{1} &= \frac{nD\rho_0^{-1/n}r^2}{K(1+n)},  \quad 
\wop{2} = r^2\left\{\frac{d}{dr}\left[\ln{\left(\frac{\rho_0r}{D}\right)}\right]\frac{d}{dr} + \frac{d^2}{dr^2}\right\},\quad
\wop{3} = -r^2\left\{\frac{H^\prime}{H}\frac{d}{dr}\left[\ln{\left(\frac{\rho_0r}{D}\right)}\right]           
                    + \left(\frac{H^\prime}{H}\right)^\prime  + \frac{H^\prime}{H}\frac{d}{dr}\right\}, \notag\\
\wop{4} &= -r^2\frac{H^\prime}{H}\frac{d}{dr}, \quad
\wop{5} = r^2\left(\frac{H^\prime}{H}\right)^2,\quad
\wop{6} = \frac{2m r\Omega}{\sbar}\frac{d}{dr}\left[\ln{\left(\frac{\rho_0\Omega}{D}\right)}\right],\notag\\
\wop{7} &= -\frac{2m r\Omega}{\sbar}\left( \frac{H^\prime}{H}\right),\quad
\wop{8} = -m^2, \quad
\wop{9} = -\frac{r^2D}{\sbar^2H^2}.
\end{align}
We have used the midplane density $\rho_0$, but it is  
straight forward to express the above in terms of $\Sigma$ using the 
relation $\Sigma = \rho_0 H(r) I_n $. 
The form of the operators above are appropriate for numerical
computations in the range of polytropic indices considered in this
paper ($n\geq1$ or $\lambda\geq0.5$). Numerical issues may arise for
smaller indices because of the $(l+ \lambda - 2)^{-1}$ factor in
$B_l$. For example, if $\lambda=0$ ($n=0.5$) and $l=2$ this factor
diverges. However, for $n=0.5$ it is more natural to use 
Chebyshev polynomials of the first kind ($T_l$) for expansion in
$\hat{z}$.  
We have performed calculations with $n=0.5$ using $T_l$, and found similar results to
those presented here.


\section{Supplementary calculations}\label{supp_calc}

\subsection{Improved simulations}\label{improved}
During the finishing stages of this paper, \cite{meheut12b} published
new simulations of the 3D RWI with improved numerical
resolution. This simulation developed 
a $m=5$ mode with growth rate $|\gamma|=0.17\Omega_0$, with
upwards motion at anti-cyclonic vortex centers and downwards motion at 
cyclonic vortex centers, which are consistent with our 
fiducial polytropic disks (\S\ref{polytropic}). 

\begin{figure}
   \centering
   \includegraphics[scale=0.42,clip=true,trim=0cm 0.1cm 0cm
     0cm]{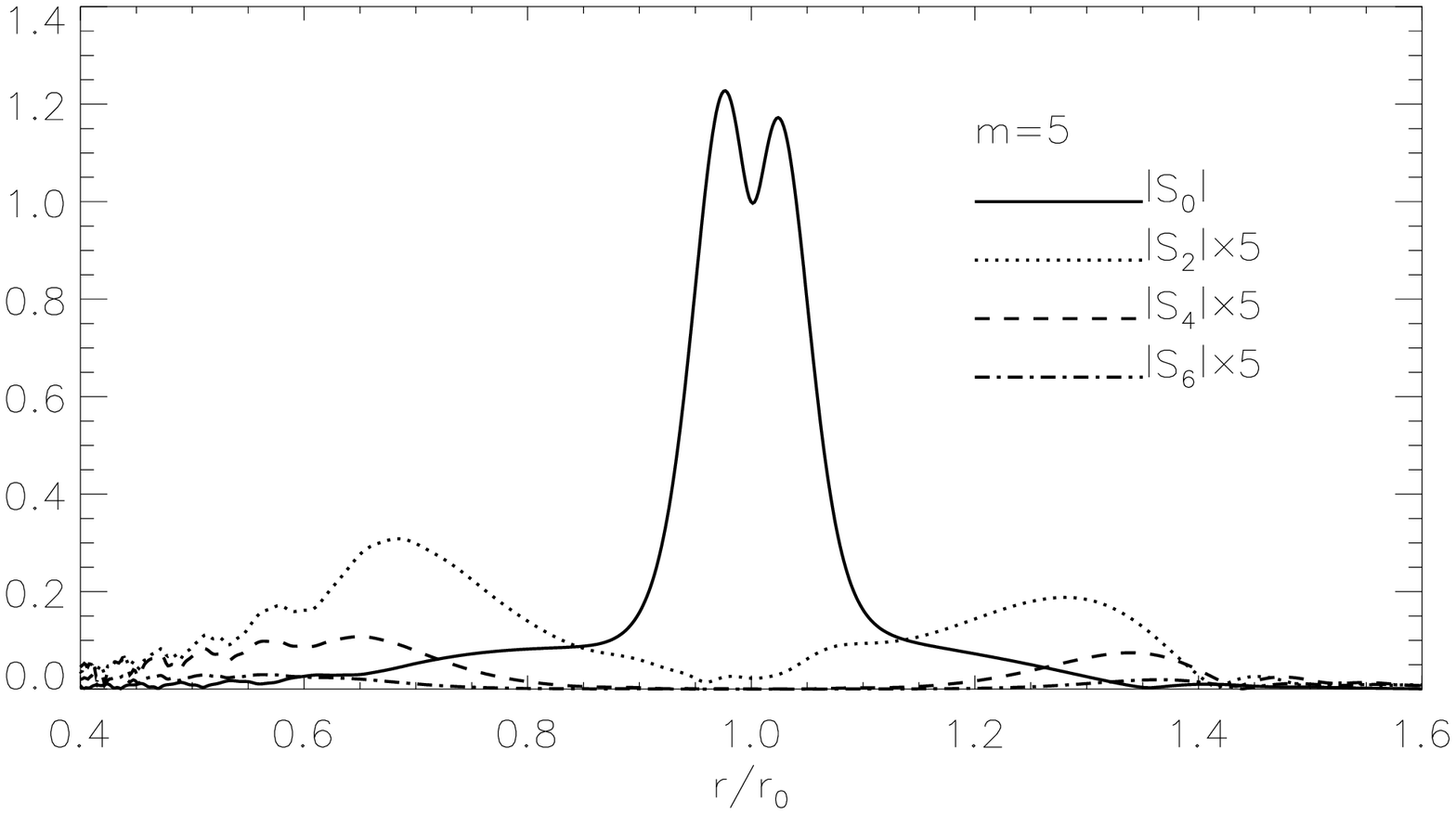}\includegraphics[scale=0.46,clip=true,trim=0.0cm 0.5cm 0.0cm
     1.0cm]{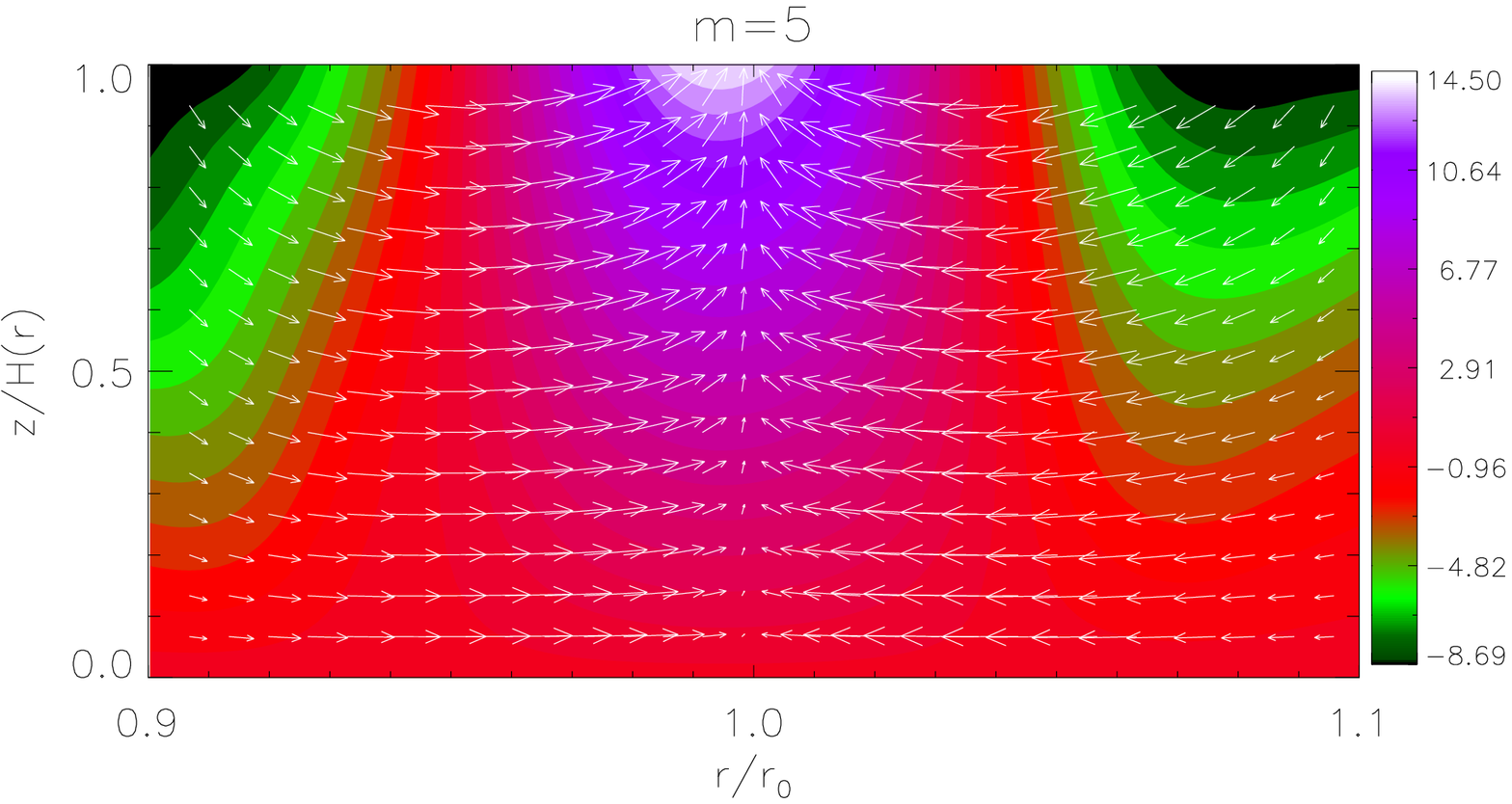} 
   \caption{ A linear $m=5$ mode found in \cite{meheut12b}'s
       $n=1.5$ polytropic disk
       model, but with a smaller bump amplitude than their
       simulation. Left: radial eigenfunctions $S_l$ normalized by
       $|S_0(r_0)|$. Right: vertical flow structure. 
     \label{meheut12b}}
 \end{figure}

We were able to find a $m=5$ linear mode provided the bump amplitude
$\amp$ in the midplane density was chosen to ensure
$\kappa^2>0$. Using $\amp=1.7$, we find $|\gamma|=0.18\Omega_0$ for
$m=5$. This mode is shown in Fig. \ref{meheut12b}. Note that $S_0$ is
still localized about $r_0$, despite the higher $m$ than those
considered in our fiducial calculations (which gave more global 
disturbances). This is because here the vortensity minimum is deep,  
with $\mathrm{min}(\kappa^2/\Omega_k^2)\simeq0.1$, so even high $m$
modes can be localized. The vertical flow at the vortex core is
upwards, as found previously.   

\cite{meheut12b} actually employed $\amp=2$, giving
$\kappa_0^2\simeq-0.2\Omega_k^2$, for which we were
unable to find a linear mode with similar growth rate as their
simulation. As $\kappa_0^2$ is more negative in their new simulation than
in \cite{meheut10}, one possibility is that an axisymmetric
disturbance  develops early on, rendering $\kappa^2\gtrsim0$ then the 
usual RWI follows. For $\amp=1.7$ we find linear growth rates
peak at $m=8$ with $|\gamma|=0.21\Omega_0$, but this is only
marginally larger than $m=5$. Differences in 
the linear and  nonlinear calculations, such as the treatment of
vertical boundaries, may then account for observation of $m=5$ in the
simulations.   


\subsection{Consistency check}\label{consistency}

We describe calculations to check the consistency between 
locally isothermal and polytropic disks and against the globally
isothermal disk presented in \cite{meheut12}. 

\begin{figure}
  \centering
  \includegraphics[width = 0.5\linewidth,clip=true,trim=0.0cm 0.5cm
    0.0cm 1.0cm]{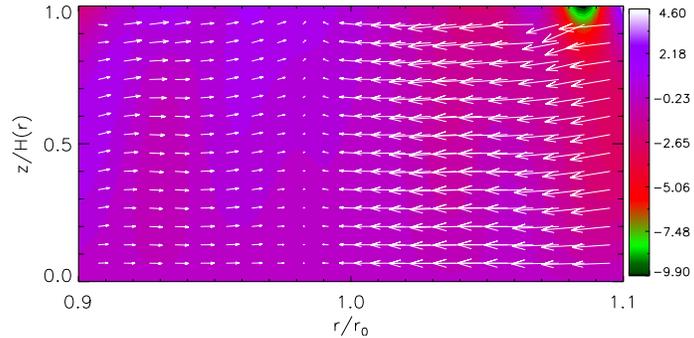} 
  \caption{
      A $m=3$ mode in the standard $n=8$ polytropic disk (growth rate
      $|\gamma|=0.055\Omega_0$). Contours of the real vertical
      velocity perturbation are shown. Arrows are the perturbed
      velocity field projected onto this plane. This figure is similar to locally
      isothermal disks in that there is very little vertical velocity
      near the vortex core, and is unlike polytropic disks with
      smaller $n$ (e.g  Fig. \ref{poly_compare_n_dvz}, which  
      shows significant upwards motion near $r=r_0$).
    \label{n8}}
\end{figure}

Noting that an isothermal disk is a special case of a polytropic disk
in the limit of large $n$, we performed a polytropic disk calculation
with $n=8,\,\amp=2.0, \, h=0.2$. Fig. \ref{n8} show that in 
this case vertical motion is much smaller than the  
horizontal flow in the co-rotation region, compared to smaller values 
of $n$ discussed in \S\ref{varn}. This is consistent with our
typical results for locally isothermal disks were the vertical velocity
vanishes at the vortex core. 

\begin{figure}
  \centering
  \includegraphics[width = 0.5\linewidth,clip=true,trim=0.0cm 0.5cm
    0.0cm 1.0cm]{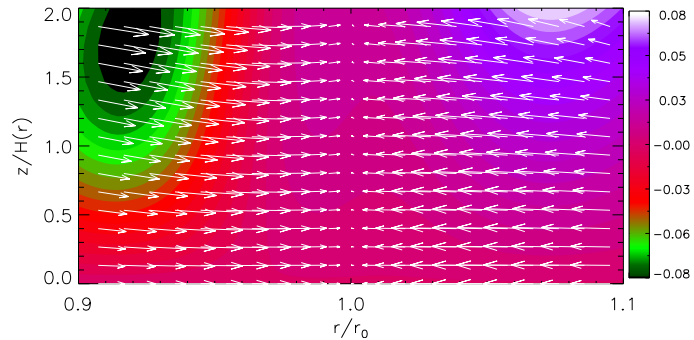} 
  \caption{
      A linear mode in the locally isothermal disk with the
      same parameter values as the \cite{meheut12}'s
      globally isothermal disk ($m=4,\,h=0.1,$ other parameters are
      the same as our fiducial case in \S\ref{isothermal}). Contours
      of the real vertical velocity perturbation is shown. Arrows are the perturbed
      velocity field projected onto this plane. The growth
      rate $|\gamma|=0.20\Omega_0$ is similar to
      \citeauthor{meheut12}.  The vertical flow is also consistent
      with their Fig. 3d, namely the vertical velocity vanishes
      near $r=r_0$. \label{meheut12_lin}}
\end{figure}

\cite{meheut12} solved the linear problem for globally isothermal
disks. Their basic state with $\Omega=\Omega(r)$ satisfy exact
radial momentum balance but adopting such a profile for locally
isothermal disks is only approximate.  
We have performed a locally isothermal calculation with 
the same parameters as \cite{meheut12}. The result is shown in
Fig. \ref{meheut12_lin}. It shares the same vertical flow implied by
\cite{meheut12}'s Fig. 3d around a maximum in the (real) density
perturbation:  
$\delta v_z>0$ near $r=1.1$, $\dd v_z<0$ near
$r=0.9$ and $\dd v_z\sim 0$ at $r=r_0$.  This suggests that a node 
in the vertical velocity at the vortex core is a generic feature for
linear RWI modes in isothermal disks. A global temperature profile
does not affect the 3D RWI significantly.  

\begin{thebibliography}{}

\bibitem[{{Abramowitz} \& {Stegun}(1965)}]{stegun65}
{Abramowitz}, M., \& {Stegun}, I.~A. 1965, {Handbook of mathematical functions
  with formulas, graphs, and mathematical tables}, ed. {Abramowitz, M.~\&
  Stegun, I.~A.}

\bibitem[{{Armitage}(2011)}]{armitage11}
{Armitage}, P.~J. 2011, \araa, 49, 195

\bibitem[{{Balbus} \& {Hawley}(1991)}]{balbus91}
{Balbus}, S.~A., \& {Hawley}, J.~F. 1991, \apj, 376, 214

\bibitem[{{Barge} \& {Sommeria}(1995)}]{barge95}
{Barge}, P., \& {Sommeria}, J. 1995, \aap, 295, L1

\bibitem[{{Chandrasekhar}(1961)}]{chandrasekhar61}
{Chandrasekhar}, S. 1961, {Hydrodynamic and hydromagnetic stability}, ed.
  {Chandrasekhar, S.}

\bibitem[{{Crespe} {et~al.}(2011){Crespe}, {Gonzalez}, \& {Arena}}]{crespe11}
{Crespe}, E., {Gonzalez}, J.-F., \& {Arena}, S.~E. 2011, in SF2A-2011:
  Proceedings of the Annual meeting of the French Society of Astronomy and
  Astrophysics, ed. {G.~Alecian, K.~Belkacem, R.~Samadi, \& D.~Valls-Gabaud},
  469--473

\bibitem[{{de Val-Borro} {et~al.}(2007){de Val-Borro}, {Artymowicz},
  {D'Angelo}, \& {Peplinski}}]{valborro07}
{de Val-Borro}, M., {Artymowicz}, P., {D'Angelo}, G., \& {Peplinski}, A. 2007,
  \aap, 471, 1043

\bibitem[{{Dong} {et~al.}(2011){Dong}, {Rafikov}, \& {Stone}}]{dong11}
{Dong}, R., {Rafikov}, R.~R., \& {Stone}, J.~M. 2011, \apj, 741, 57

\bibitem[{{Gammie}(1996)}]{gammie96}
{Gammie}, C.~F. 1996, \apj, 457, 355

\bibitem[{{Goldreich} {et~al.}(1986){Goldreich}, {Goodman}, \&
  {Narayan}}]{goldreich86}
{Goldreich}, P., {Goodman}, J., \& {Narayan}, R. 1986, \mnras, 221, 339

\bibitem[{{Goldreich} \& {Tremaine}(1979)}]{goldreich79}
{Goldreich}, P., \& {Tremaine}, S. 1979, \apj, 233, 857

\bibitem[{{Goldreich} \& {Tremaine}(1980)}]{goldreich80}
---. 1980, \apj, 241, 425

\bibitem[{{Knobloch} \& {Spruit}(1986)}]{knobloch86}
{Knobloch}, E., \& {Spruit}, H.~C. 1986, \aap, 166, 359

\bibitem[{{Koller} {et~al.}(2003){Koller}, {Li}, \& {Lin}}]{koller03}
{Koller}, J., {Li}, H., \& {Lin}, D.~N.~C. 2003, \apjl, 596, L91

\bibitem[{{Lesur} \& {Papaloizou}(2009)}]{lesur09}
{Lesur}, G., \& {Papaloizou}, J.~C.~B. 2009, \aap, 498, 1

\bibitem[{{Lesur} \& {Papaloizou}(2010)}]{lesur10}
---. 2010, \aap, 513, A60

\bibitem[{{Li} {et~al.}(2001){Li}, {Colgate}, {Wendroff}, \& {Liska}}]{li01}
{Li}, H., {Colgate}, S.~A., {Wendroff}, B., \& {Liska}, R. 2001, \apj, 551, 874

\bibitem[{{Li} {et~al.}(2000){Li}, {Finn}, {Lovelace}, \& {Colgate}}]{li00}
{Li}, H., {Finn}, J.~M., {Lovelace}, R.~V.~E., \& {Colgate}, S.~A. 2000, \apj,
  533, 1023

\bibitem[{{Li} {et~al.}(2005){Li}, {Li}, {Koller}, {Wendroff}, {Liska},
  {Orban}, {Liang}, \& {Lin}}]{li05}
{Li}, H., {Li}, S., {Koller}, J., {Wendroff}, B.~B., {Liska}, R., {Orban},
  C.~M., {Liang}, E.~P.~T., \& {Lin}, D.~N.~C. 2005, \apj, 624, 1003

\bibitem[{{Li} {et~al.}(2009){Li}, {Lubow}, {Li}, \& {Lin}}]{li09}
{Li}, H., {Lubow}, S.~H., {Li}, S., \& {Lin}, D.~N.~C. 2009, \apjl, 690, L52

\bibitem[{{Li} {et~al.}(2003){Li}, {Goodman}, \& {Narayan}}]{li03}
{Li}, L.-X., {Goodman}, J., \& {Narayan}, R. 2003, \apj, 593, 980

\bibitem[{{Lin} \& {Papaloizou}(1986)}]{lin86}
{Lin}, D.~N.~C., \& {Papaloizou}, J. 1986, \apj, 309, 846

\bibitem[{{Lin} \& {Papaloizou}(2010)}]{lin10}
{Lin}, M.-K., \& {Papaloizou}, J.~C.~B. 2010, \mnras, 405, 1473

\bibitem[{{Lin} \& {Papaloizou}(2011{\natexlab{a}})}]{lin11a}
---. 2011{\natexlab{a}}, \mnras, 415, 1426

\bibitem[{{Lin} \& {Papaloizou}(2011{\natexlab{b}})}]{lin11b}
---. 2011{\natexlab{b}}, \mnras, 415, 1445

\bibitem[{{Lovelace} {et~al.}(1999){Lovelace}, {Li}, {Colgate}, \&
  {Nelson}}]{lovelace99}
{Lovelace}, R.~V.~E., {Li}, H., {Colgate}, S.~A., \& {Nelson}, A.~F. 1999,
  \apj, 513, 805

\bibitem[{{Lyra} {et~al.}(2008){Lyra}, {Johansen}, {Klahr}, \&
  {Piskunov}}]{lyra08}
{Lyra}, W., {Johansen}, A., {Klahr}, H., \& {Piskunov}, N. 2008, \aap, 491, L41

\bibitem[{{Lyra} {et~al.}(2009){Lyra}, {Johansen}, {Zsom}, {Klahr}, \&
  {Piskunov}}]{lyra09}
{Lyra}, W., {Johansen}, A., {Zsom}, A., {Klahr}, H., \& {Piskunov}, N. 2009,
  \aap, 497, 869

\bibitem[{{Meheut} {et~al.}(2010){Meheut}, {Casse}, {Varniere}, \&
  {Tagger}}]{meheut10}
{Meheut}, H., {Casse}, F., {Varniere}, P., \& {Tagger}, M. 2010, \aap, 516, A31

\bibitem[{{Meheut} {et~al.}(2012{\natexlab{a}}){Meheut}, {Keppens}, {Casse}, \&
  {Benz}}]{meheut12b}
{Meheut}, H., {Keppens}, R., {Casse}, F., \& {Benz}, W. 2012{\natexlab{a}},
  ArXiv e-prints

\bibitem[{{Meheut} {et~al.}(2011{\natexlab{a}}){Meheut}, {Varniere}, \&
  {Benz}}]{meheut11a}
{Meheut}, H., {Varniere}, P., \& {Benz}, W. 2011{\natexlab{a}}, in EPSC-DPS
  Joint Meeting 2011, held 2-7 October 2011 in Nantes, France, 1054

\bibitem[{{Meheut} {et~al.}(2011{\natexlab{b}}){Meheut}, {Varniere}, {Casse},
  \& {Tagger}}]{meheut11b}
{Meheut}, H., {Varniere}, P., {Casse}, F., \& {Tagger}, M. 2011{\natexlab{b}},
  in EPSC-DPS Joint Meeting 2011, held 2-7 October 2011 in Nantes, France, 1059

\bibitem[{{Meheut} {et~al.}(2012{\natexlab{b}}){Meheut}, {Yu}, \&
  {Lai}}]{meheut12}
{Meheut}, H., {Yu}, C., \& {Lai}, D. 2012{\natexlab{b}}, \mnras, 2748

\bibitem[{{Muto} {et~al.}(2010){Muto}, {Suzuki}, \& {Inutsuka}}]{muto10}
{Muto}, T., {Suzuki}, T.~K., \& {Inutsuka}, S.-i. 2010, \apj, 724, 448

\bibitem[{{Narayan} {et~al.}(1987){Narayan}, {Goldreich}, \&
  {Goodman}}]{narayan87}
{Narayan}, R., {Goldreich}, P., \& {Goodman}, J. 1987, \mnras, 228, 1

\bibitem[{{Okazaki} \& {Kato}(1985)}]{okazaki85}
{Okazaki}, A.~T., \& {Kato}, S. 1985, \pasj, 37, 683

\bibitem[{{Ou} {et~al.}(2007){Ou}, {Ji}, {Liu}, \& {Peng}}]{ou07}
{Ou}, S., {Ji}, J., {Liu}, L., \& {Peng}, X. 2007, \apj, 667, 1220

\bibitem[{{Papaloizou} \& {Pringle}(1984)}]{papaloizou84}
{Papaloizou}, J.~C.~B., \& {Pringle}, J.~E. 1984, \mnras, 208, 721

\bibitem[{{Papaloizou} \& {Pringle}(1985)}]{papaloizou85}
---. 1985, \mnras, 213, 799

\bibitem[{{Papaloizou} \& {Pringle}(1987)}]{papaloizou87}
---. 1987, \mnras, 225, 267

\bibitem[{{Reg{\'a}ly} {et~al.}(2012){Reg{\'a}ly}, {Juh{\'a}sz}, {S{\'a}ndor},
  \& {Dullemond}}]{regaly12}
{Reg{\'a}ly}, Z., {Juh{\'a}sz}, A., {S{\'a}ndor}, Z., \& {Dullemond}, C.~P.
  2012, \mnras, 419, 1701

\bibitem[{{Takeuchi} \& {Miyama}(1998)}]{takeuchi98}
{Takeuchi}, T., \& {Miyama}, S.~M. 1998, \pasj, 50, 141

\bibitem[{{Tanaka} {et~al.}(2002){Tanaka}, {Takeuchi}, \& {Ward}}]{tanaka02}
{Tanaka}, H., {Takeuchi}, T., \& {Ward}, W.~R. 2002, \apj, 565, 1257

\bibitem[{{Terquem}(2008)}]{terquem08}
{Terquem}, C.~E.~J.~M.~L.~J. 2008, \apj, 689, 532

\bibitem[{{Umurhan}(2008)}]{umurhan08}
{Umurhan}, O.~M. 2008, \aap, 489, 953

\bibitem[{{Umurhan}(2010)}]{umurhan10}
---. 2010, \aap, 521, A25

\bibitem[{{Umurhan}(2012)}]{umurhan12}
---. 2012, ArXiv e-prints

\bibitem[{{Varni{\`e}re} \& {Tagger}(2006)}]{varniere06}
{Varni{\`e}re}, P., \& {Tagger}, M. 2006, \aap, 446, L13

\bibitem[{{Yang} \& {Menou}(2010)}]{yang10}
{Yang}, C.-C., \& {Menou}, K. 2010, \mnras, 402, 2436

\bibitem[{{Yu} \& {Li}(2009)}]{yu09}
{Yu}, C., \& {Li}, H. 2009, \apj, 702, 75

\bibitem[{{Yu} {et~al.}(2010){Yu}, {Li}, {Li}, {Lubow}, \& {Lin}}]{yu10}
{Yu}, C., {Li}, H., {Li}, S., {Lubow}, S.~H., \& {Lin}, D.~N.~C. 2010, \apj,
  712, 198

\bibitem[{{Zhang} \& {Lai}(2006)}]{zhang06}
{Zhang}, H., \& {Lai}, D. 2006, \mnras, 368, 917

\end{thebibliography}
\end{document}